# A Double Machine Learning Approach to Estimate the Effects of Musical Practice on Student's Skills

Michael C. Knaus[*]

**Abstract:** This study investigates the dose-response effects of making music on youth development. Identification is based on the conditional independence assumption and estimation is implemented using a recent double machine learning estimator. The study proposes solutions to two highly practically relevant questions that arise for these new methods: (i) How to investigate sensitivity of estimates to tuning parameter choices in the machine learning part? (ii) How to assess covariate balancing in high-dimensional settings? The results show that improvements in objectively measured cognitive skills require at least medium intensity, while improvements in school grades are already observed for low intensity of practice.

---

[*] University of St. Gallen. Michael C. Knaus is also affiliated with IZA, Bonn. Financial support from the Swiss National Science Foundation (SNSF) for the project "Causal Analysis with Big Data" (grant number SNSF 407540_166999), which is part of the Swiss National Research Programme "Big Data" (NRP 75) is gratefully acknowledged. This paper uses data from the National Educational Panel Study (NEPS): Starting Cohort Grade 9, doi:10.5157/NEPS:SC4:7.0.0. From 2008 to 2013, NEPS data was collected as part of the Framework Program for the Promotion of Empirical Educational Research funded by the German Federal Ministry of Education and Research (BMBF). As of 2014, NEPS is carried out by the Leibniz Institute for Educational Trajectories (LIfBi) at the University of Bamberg in cooperation with a nationwide network. Chris Hansen, Michael Lechner, Jeff Smith, Martin Spindler, Stefan Wager and Michael Zimmert are greatly acknowledged for their helpful comments and suggestions. The usual disclaimer applies.

# 1 Introduction

The combination of so-called machine learning and causal inference is currently an active area of methodological research. Especially new methods for the estimation of average treatment effects (e.g., Athey, Imbens, & Wager, 2018; Chernozhukov, et al., 2018; Farrell, 2015) and the estimation of heterogeneous treatment effects (e.g., Athey, Tibshirani, & Wager, 2018; Wager & Athey, 2018) are recently proposed. Few papers start to put causal machine learning estimators for heterogeneous treatment effects into practice and discuss practically relevant issues (e.g., Bertrand, Crépon, Marguerie, & Premand, 2017; Davis & Heller, 2017; Knaus, Lechner, & Strittmatter, 2017, 2018). However, applications that do the same for average treatment effects are currently missing. Still, these methods have the potential to improve causal analysis in observational studies. This paper studies the effects of musical practice on child development in an observational study and provides novel ideas to fruitfully combine these methods and standard empirical practices.

The analysis is motivated by the relevance of cognitive and non-cognitive skills for success at school and in the labor market (Kautz, Heckman, Diris, Weel, & Borghans, 2014). Development of these skills is therefore of fundamental individual, economic and societal importance. Besides schools and families as the main drivers of human capital accumulation of children, the economic literature on child development shows a recent interest in understanding the role of extracurricular activities like sports or music (Cabane, Hille, & Lechner, 2016; Felfe, Lechner, & Steinmayr, 2016; Hille & Schupp, 2015). Previous evidence suggests that engagement in these extracurricular activities *per se* has positive effects on at least some measureable cognitive and non-cognitive skills.[1] However, evidence with respect to the intensity of playing music is missing so far.[2] Shedding light on the dose-response relation of playing music has important implications for individuals, parents and policymakers. It allows to answer at least two important questions: (i) Which level of engagement is required to generate the observed gains? (ii) Is more always better or does very intense musical practice harm human capital accumulation by crowding out other productive activities?

---

[1] These findings are in line with results in neuroscience and sociology (e.g., Bergman Nutley et al., 2014; Eccles et al., 2003).

[2] Cabane et al. (2016) is the only study that accounts for the intensity of activities in some way by distinguishing between sports and competitive sports.

The effects in this and in previous studies are identified using the conditional independence assumption (CIA) that demands usually a large set of control variables to be plausible. Such analyses are a workhorse for empirical researchers to identify causal parameters like average treatment effects in observational studies.[3] The estimation of these parameters usually requires to select the controls that enter the analysis and their functional form (see for reviews, Abadie & Cattaneo, 2018; Imbens & Wooldridge, 2009). A flexible set of potential controls with polynomials and interactions easily leads to a setting where the number of potential controls exceeds the number of observations. Standard methods are not feasible in such high-dimensional settings and require some more or less principled variable selection. Double machine learning (DML) shows how the estimation of causal effects can be split into several prediction problems (Chernozhukov, Chetverikov, et al. 2018). Thus, it allows to leverage methods from the machine learning literature that are developed for high-dimensional prediction problems (see for an overview, e.g., Hastie, Tibshirani, & Friedman, 2009). DML enables the integration of these methods into causal analysis with observational data and to control for selection bias in an objective and data-driven way.[4]

This paper contributes to two strands of literature. First, it adds to the literature about extracurricular activities and youth development by investigating a potential dose-response relation between musical practice and cognitive and non-cognitive skills. To this end, the German National Economic Panel Study (NEPS) (Blossfeld, Roßbach, & von Maurice, 2011) provides unique information for all dimensions of the analysis. Besides measures of music and outcomes of interest, the detailed parental information regarding cultural preferences in the NEPS data allow a more credible identification of causal effects compared to previous studies. Second, the paper contributes to the young causal machine learning literature. While the idea of DML triggered a variety of methodological contributions,[5] applications that are not run for expository purposes in these contributions are missing so far. This paper provides a first step to combine applied empirical practices and standards with these new methods. Specifically, it builds on the DML estimator of Farrell (2015) for average treatment effects with multivalued treatments. In the absence of any established procedures for applications, the paper addresses two practically important questions: (i) How can we check covariate balancing of the estimator, which is standard for estimators based on the propensity score (see, e.g., Lee, 2013)? This paper derives a weighted representation of the DML method that can be used to assess covariate balancing with established measures. (ii) How can we assess the sensitivity of our estimators to tuning parameter choices? These tuning parameters are at the core of any machine learning algorithm and control model complexity. Out-of-sample prediction quality does heavily depend on their choice. The same might be suspected when using these predictions for causal inference. This paper proposes a data-driven assessment that is inspired by the one standard error rule (1SE) of Breiman, Friedman, Stone and Olshen (1984).

---

[3] This assumption is also known as unconfoundedness, exogeneity or selection on observables assumption.
[4] The overview of Athey (2018) discusses this point and other uses of machine learning for economic analysis.

The results show that statistically significant improvements in objectively measured cognitive skills require at least medium intensity of practice, which means 8 to 22 days making music per month. However, improvements in school grades are already observed for low intensity practice with at least one day engaging in musical activities per month. Using the Big Five as a measure of non-cognitive skills, we find significant improvements of agreeableness and openness. On the methodological side, DML successfully balances a high-dimensional set of covariates by including only a low-dimensional set of controls in this application. Furthermore, the empirical results are robust to different choices of methods and tuning parameters in the machine learning part.

The paper proceeds as follows. The next section provides a brief overview of the previous literature on model selection for causal analysis and the literature leading to DML. Section 3 describes the NEPS data. Section 4 discusses identification via CIA, the chosen DML estimator and the innovations of this paper. Section 5 provides the results and section 6 concludes. Appendices A to E provide additional material. The accompanying R package `dmlmt` for DML with multivalued treatments and an illustrative example building on Chernozhukov, Hansen and Spindler (2016) are provided at https://github.com/mcknaus/dmlmt.

---

[5] Recent examples being Antonelli and Dominici (2018), Athey and Wager (2017), Chernozhukov, Goldman, Semenova, and Taddy (2017), Chernozhukov, Newey, and Robins (2018), Luo and Spindler (2017) and Mackey, Syrgkanis and Zadik (2017).

## 2 Literature review

The DML methodology to estimate average treatment effects is a rather recent development. This section gives a brief overview of the related literature and refers the interested reader to the original papers for the technical details.

Most practically relevant estimators based on the CIA require to specify a model for the conditional expectation of the outcome, for the conditional treatment probability (propensity score) or for both. However, the literature provides little guidance for researchers on how to conduct proper model selection. This is problematic because the number of potential variables can easily exceed the number of observations if researchers include interactions and polynomials of the base variables. Hirano and Imbens (2001) propose a systematic way of model selection by keeping only variables that are statistically significant at a pre-determined level in the outcome or propensity score model. However, this is not feasible for a high-dimensional set of potential controls. An alternative for propensity score based methods proceeds by iteratively adding interaction terms and polynomials to the propensity score model until the covariate distributions in treatment and control groups are considered as balanced (Dehejia & Wahba, 1999, 2002; Rosenbaum & Rubin, 1984). Other approaches apply machine learning techniques to flexibly estimate the propensity score (B. K. Lee, Lessler, & Stuart, 2010; McCaffrey, Ridgeway, & Moral, 2004; Wyss et al., 2014). These methods conduct standard statistical inference that ignores the model selection step.

All the reviewed approaches can be problematic for two related reasons. First, Leeb and Pötscher (2005, 2008) show that such "post-model-selection estimators" might lead to invalid statistical inference. They note that inference procedures after model selection are not uniformly consistent. However, uniform consistency is necessary to use asymptotic properties of estimators as approximations in finite samples. As a consequence, statistical inference that ignores the model selection step in finite samples can be misleading. The second problem arises when either only the outcome or only the propensity score model is considered in the model selection step. Belloni, Chernozhukov and Hansen (2014a, 2014b) illustrate how these "single-equation approaches" can fail to provide valid statistical inference. This problem arises because the CIA requires to control for variables that affect the treatment probability *and* the outcome. Model selection that is only based on one of the two might miss variables that have a small coefficient in the considered but a large coefficient in the other model. As a consequence, single equation approaches might fail to find relevant controls and can be biased.

Belloni et al. (2014b) and Farrell (2015) offer a constructive solution to both problems. Their approaches build on Hahn's (1998) efficient score for semiparametric average treatment effect estimation.[6] In this setting, the conditional expectations of the outcome and the propensity score serve as potentially high-dimensional nuisance parameter. The goal is then to use machine learning tools that provide high-quality approximations of these nuisance parameters. The combination of efficient score and high-quality prediction methods allows Belloni et al. (2014b) and Farrell (2015) to provide uniformly valid inference also after model selection. This is achieved by considering model selection as high-quality approximation of all nuisance parameters instead of perfect variable selection in either outcome or propensity score model.

Belloni, Chernozhukov, Fernández-Val, and Hansen (2017) generalize these ideas to all parameters that are identified via moment conditions that satisfy Neyman orthogonality (Neyman, 1959). Such moment conditions are immune to small errors in the approximation of the nuisance parameters. Chernozhukov, Chetverikov, et al. (2018, 2017) call this approach DML and discuss how a variety of machine learning algorithms can be applied for causal inference in this framework. However, applications using these methods are missing so far and practical issues still need to be investigated.

---
[6] For a detailed discussion of the connections to semiparametric theory, see Zimmert (2018).

# 3 Data

## 3.1 National Educational Panel Study

The empirical analysis is based on the German National Educational Panel Study (NEPS) (Blossfeld et al., 2011). Specifically, we use the first wave of starting cohort four, which was conducted in autumn 2010 with 15,577 students in the 9th grade. The student survey and tests were performed in classrooms. Afterwards, parents were surveyed in telephone interviews. The cohort of 9th graders is particularly well-suited for the research question at hand because they were exclusively and extensively asked about their extracurricular activities including intensity of musical practice (Frahm et al., 2011). The number of observations available for the analysis is 6,898. Appendix A.1 provides details about the sample preparation.

The *outcome variables* in this analysis can be divided into cognitive and non-cognitive skills. The measures of domain-specific cognitive skills are obtained from standardized tests in math, reading literacy, information and communication technology (ICT) literacy, and language proficiency (vocabulary test) (Artelt, Weinert, & Carstensen, 2013). Further, self-reported German and math grades can be used to assess whether potential differences in the objectively measured skills are also mirrored in the more subjective evaluation by teachers. Non-cognitive skills are assessed by using the Big Five measures of personality traits: extraversion, agreeableness, conscientiousness, neuroticism, and openness (McCrae & Costa, 1999).

The strategy based on the CIA described below requires a large set of background characteristics to serve as *control variables*. The NEPS provides very detailed information in the individual and parental questionnaire from where we extract 377 potential base control variables.[7]

---

[7] A detailed description of the considered variables and how they are coded is provided in Table A.1.2 of Appendix A.1.

They contain individual characteristics, parental preferences for leisure activities, parental work, household economic conditions, parenting attitudes, household demographics, home possessions, information about social circle of parents, and regional information.

## 3.2 Measurement of music intensity

Cohort four of the NEPS data provides a unique measure for the intensity of musical activities. Students are asked the following question: *"On how many days in the last month have you made music, e.g. played an instrument or sung in a choir? Making music on the computer does not count. On about …"*. The number of days that are reported serve as our measure of intensity. Figure 1 shows the distribution of the answers for all students (left) and those that report a positive number of music days (right). The left graph shows that the majority (52%) reports no musical practice in the previous month.

*Figure 1: Distribution of reported days with musical practice in month before interview*

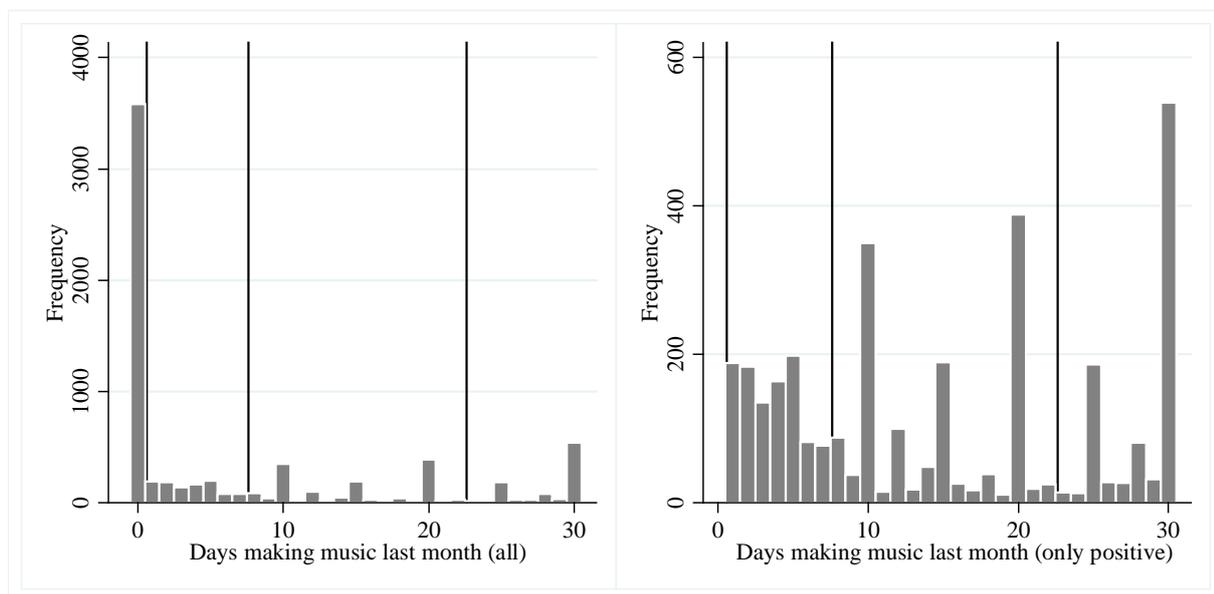

Note: The left graph shows the frequencies in the full sample and the right graph only for those reporting at least one day of musical practice in the last month.

One obvious feature of the intensity measure is the rounding pattern at steps of five and especially at steps of ten. Such rounding is frequently observed in surveys when people are asked to recall past frequencies. The analysis is therefore based on discretized intensity measures because the continuous measure might be affected by systematic measurement error and because precision for values between the peaks would be low due to very few observations for these values. The intensity measure is split into four categories for student $i$ depending on the reported $days/month_i$:[8]

$$Intensity_i \begin{cases} No_i & \text{if } days/month_i = 0 \\ Low_i & \text{if } days/month_i \in [1,7] \\ Med_i & \text{if } days/month_i \in [8,22] \\ High_i & \text{if } days/month_i > 22 \end{cases} \quad (1)$$

Further, we define a binary indicator $Any_i$ being one for any positive intensity. As such binary indicators were used in previous studies, we want to check how our findings compare to their results based on a binary indicator before investigating different intensities.[9]

## 4 Econometric approach

### 4.1 Identification

Identification of the effects of music is complicated by the fact that the decision to play music and the intensity are not made at random. A lot of background characteristics like socio-economic status could influence the decision to play music and the outcomes of interest simultaneously. These so-called confounders need to be controlled for if we are interested in causal effects.

To fix ideas, consider the potential outcome framework of Rubin (1974) in a multivalued treatment setting with $T + 1$ different treatments $D_i \in \{0,1,...,T\}$ (Imbens, 2000; Lechner, 2001). Random variables are indicated by capital letters and the realizations of these random variables by lowercase letters. Each individual $i = 1,...,n$ has a potential outcome $Y_i^t$ for each value of the treatment $D_i = t$ but only the potential outcome of the realized treatment value is observed. The observed outcome is therefore $Y_i = \sum_t \mathbf{1}\{D_i = t\}Y_i^t$ and the potential outcomes with $D_i \neq t$ remain latent. However, we aim at estimating the mean of the potential outcomes $\mu_t = E[Y_i^t]$ and their differences. For example, the average treatment effect (ATE), $\mu_1 - \mu_0 = E[Y_i^1 - Y_i^0]$, is often a parameter of interest in the case of a binary treatment variable $D_i \in \{0,1\}$. The multivalued treatment setting considered here provides a larger set of average treatment effects by allowing any possible pairwise comparison $\gamma_{m,k} = \mu_m - \mu_k = E[Y_i^m - Y_i^k]$ for $m \neq k$.[10]

---

[8] Also indicated by the black lines in Figure 1. The splits are placed in the middle of steps of five assuming that students round to the next step of five. The active students are divided in three categories of similar size to facilitate the interpretation of the estimated parameters. Specifications with more or alternative categories are very similar to those presented later but the large number of estimated parameters makes the interpretation very cumbersome.

[9] Appendix A.2 provides a detailed discussion about what playing music and the different intensities means and how these students spend their time otherwise.

The selection into different treatment levels in observational studies leads to $\mu_t \neq E[Y_i]$ due to selection bias. However, identification of $\mu_t$ can still be achieved if a vector of covariates $X_i$ exists such that the following two assumptions are fulfilled:

**Assumption 1** (Conditional independence): $Y_i^t \coprod D_i | X_i = x, \forall\, t, i$ and $\forall x \in \chi$.

**Assumption 2** (Common support): $P[D_i = t | X_i = x] > 0, \forall\, t, i$ and $\forall x \in \chi$.

Both assumptions can be summarized as the strong ignorability assumption (Rosenbaum & Rubin, 1983). The first assumption means that the treatment status is as good as randomly assigned conditional on covariates $X_i$. The second assumption requires that any unit needs to have a non-zero probability to receive each of the treatments. These assumptions allow the researcher to identify the average potential outcomes $\mu_t$ and consequently the causal effects $\gamma_{m,k}$. The plausibility of the common support assumption can be assessed and seems to be unproblematic in this application as we discuss in Appendix E. The CIA is however untestable and careful arguments need to be made about the plausibility in each specific application.

The setting in this paper deals with the four treatment levels defined in equation (1). In this case, the CIA requires that we observe all variables that influence the decision to be musically active and its intensity as well as the outcomes of interest simultaneously. The decision process that leads to the observed intensity of music can be conceptualized as a three-stage process. The first two steps follow Cabane et al. (2016) and consider first the decision to engage in some extracurricular activity or not at all. In the second stage, students decide whether to play music or to engage in a different activity. If they decide to play music, the third step is to choose the intensity of musical practice.

---

[10] Note that the discussed effects are estimated for the population. In general, effects for different target populations are available like the average treatment effect on the treated and alike as discussed in detail, e.g., in Lechner (2001).

We control for the first stage of selection by considering only students that are at least active in one extracurricular activity. Thus, selection into being active is implicitly controlled for because all remaining students with intensity $No_i$ are active. This conditioning makes the CIA more plausible but also restricts the population for which the effects are identified to extracurricularly active students.

The remaining selection to be controlled for concerns the decision to engage in music and not in something else as well as the intensity decision conditional on making music. Following the extensive discussions in Hille and Schupp (2015) and Cabane et al. (2016), we control for parental tastes, parenting attitudes, economic conditions of a household, parents work characteristics, household demographics, social circle, student characteristics as well as regional and school characteristics. The discussion in Appendix A.3 spells out why this categories are required to make the CIA plausible and how they are measured in the NEPS data. It is also investigated if selection is present and which variables are the main drivers into musical practice. As suspected, we observe there a clear selection into playing music and also in the intensity of musical practice. However, the NEPS data provide a large set of variables that are needed to make identification via CIA plausible. The remaining threats to identification most likely stem from unobserved personality traits that are shown to be potentially related to musical practice (Corrigall, Schellenberg, & Misura, 2013) and the absence of better measures of early ability.

Table A.1.2 in Appendix A.1 documents how we extract 377 potential control variables from the NEPS data. One concern is that some of these variables might themselves be outcomes of musical practice and could be endogenous controls. This would violate the following assumption that is required for identification for the causal effects (see e.g., Lechner, 2008):

**Assumption 3** (Exogeneity of controls): $X_i^m = X_i^k, \forall\, m, k, i.$

The concern in this study is that musical practice measured in 9[th] grade implies usually a history of musical practice in earlier years. Several control variables also measured in 9[th] grade are prone to being influenced by this musical history. For example, measuring how often parents visit the opera or rock concerts could well be driven by the fact that they accompany their musically interested children. It is not clear *a priori* whether these endogeneity concerns are relevant for the estimation or not. The baseline specification therefore omits the 49 variables that are suspect to being endogenous.[11] One sensitivity check adds the potentially problematic variables to the analysis and finds that the results are not sensitive to their inclusion.

## 4.2 Estimation

The previous section argued that the CIA is plausible given the rich set of control variables such that average potential outcomes and average causal effects can be identified. Common estimators in this multivalued treatment setting under the CIA apply either regression adjustment based on modelling the conditional expectations of the outcome, (generalized) propensity score matching or weighting based on modelling the conditional treatment probability, or a combination of both (see for recent overviews, Linden, Uysal, Ryan, & Adams, 2016; Yang et al., 2016). Another possibility is to exploit the ordered nature of the treatments and to match on the linear index of ordinal non-linear models (Joffe & Rosenbaum, 1999; Lu, Zanutto, Hornik, & Rosenbaum, 2001). All these options involve a decision about the set of confounders that are used to model either the outcome, the treatment, or both. Further, the functional form in which these variables enter the analysis has to be chosen.

While the reasoning in the previous section and Appendix A.3 is helpful to determine the potentially important groups of control variables, no such reasoning is available to choose the set of variables that should finally enter the analysis. Many of the controls are highly correlated as they measure socio-economic status from different angles. Including all of them might lead to overfitting and substantial efficiency losses. However, even if we could identify the relevant controls or include all of them, no theory would tell us in which functional form they should enter the specification. The majority of the applied papers consider at this step that the main effects enter linearly. This assumes implicitly that interactions and non-linearities are irrelevant, which is rather restrictive given that there is usually no theoretical justification for such an assumption. If we are not willing to impose such restrictions on the functional form and allow, e.g., for second order interactions and up to 4th order polynomials for continuous variables, the number of variables that could be considered in this application already rises to about 60,000. In that case, we would end up with nearly ten times more variables than observations.

---

[11] The last column of Table A.1.2 in Appendix A.1 indicates all the variables considered as potentially endogenous.

DML can deal with such high-dimensional settings under CIA as described in the literature review above. These methods require two things: (i) high-quality predictions for the outcome and the treatment probabilities, respectively, and (ii) scores for the parameters of interest that fulfill the Neyman orthogonality condition (Neyman, 1959). Farrell (2015) develops such a DML estimator for multivalued treatments. Therefore, it is a natural candidate for the research question at hand.

The estimator proposed in Farrell (2015) is based on the efficient score for the average potential outcome under CIA (Cattaneo, 2010; Hahn, 1998; Robins, Rotnitzky, & Zhao, 1994),

$$\mu_t = E\left[\frac{d_i^t(Y_i - \mu_t(X_i))}{p_t(X_i)} + \mu_t(X_i)\right], \forall\, t \qquad (2)$$

where $d_i^t = \mathbf{1}\{D_i = t\}$, $\mu_t(x) = E[Y_i|D_i = t, X_i = x]$ and $p_t(x) = P[D_i = t|X_i = x]$ denote the treatment indicator, the conditional expectation of the outcome and the conditional probability for treatment *t*, respectively. Being a semiparametrically efficient score, equation (2) fulfills the Neyman orthogonality condition automatically (Chernozhukov, Chetverikov, et al., 2017). This means that the derivative of the moment condition (2) with respect to the so-called nuisance parameters $\mu_t(x)$ and $p_t(x)$ is equal to zero at the true value $\mu_t$. As a consequence, the score in (2) is robust to small errors in these nuisance parameters.

Farrell (2015) shows that an estimator based on sample analogues of equation (2) is square-root-n consistent and asymptotically normal if consistent estimators are used to approximate $\mu_t(x)$ and $p_t(x)$ and the product of their convergence rates reaches $n^{-1/2}$. This is fulfilled if both estimators converge at $n^{-1/4}$.

The variance in the i.i.d. setting is given by the square of the efficient score:

$$\sigma^2_{\mu,t} = E\left[\left(\frac{d_i^t(Y_i - \mu_t(X_i))}{p_t(X_i)} + \mu_t(X_i) - \mu_t\right)^2\right], \forall\, t. \qquad (3)$$

Estimates of the mean potential outcomes are achieved in three steps: (i) get predictions for the conditional outcome $\hat{\mu}_t(x)$, (ii) get predictions for the conditional treatment probability $\hat{p}_t(x)$, (iii) plug both predictions into the sample analogues of equations (2) and (3) to estimate $\hat{\mu}_t$ and $\hat{\sigma}^2_{\mu,t}$.

Pairwise treatment effects are obtained by subtracting the efficient scores for the respective potential outcomes,

$$\gamma_{m,k} = E\left[\frac{d_i^m(Y_i - \mu_m(X_i))}{p_m(X_i)} + \mu_m(X_i) - \frac{d_i^k(Y_i - \mu_k(X_i))}{p_k(X_i)} - \mu_k(X_i)\right], \forall\, m, k. \qquad (4)$$

The corresponding variance is given by

$$\sigma^2_{\gamma_{m,k}} = E\left[\left(\frac{d_i^m(Y_i - \mu_m(X_i))}{p_m(X_i)} + \mu_m(X_i) - \mu_m - \frac{d_i^k(Y_i - \mu_k(X_i))}{p_k(X_i)} - \mu_k(X_i) + \mu_k\right)^2\right], \forall m, k. \quad (5)$$

### 4.3 Implementation

We implement the Farrell (2015) estimator using Post-Lasso (Belloni & Chernozhukov, 2013) with cross-validation to choose the penalty term for prediction of the nuisance parameters. This deviates from the expository application of Farrell (2015) that applies group Lasso and asymptotic penalty terms. In this section, we describe the modified implementation before sections 4.4 and 4.5 explain how the modifications enable us to conduct standard balancing checks of the covariates and to assess the sensitivity of the analysis to the machine learning part.

The Post-Lasso is based on the Lasso estimator proposed by Tibshirani (1996).[12] The Lasso solves the following optimization problem:

$$\min_{\beta} \left[ \sum_{i=1}^{n} (Y_i - X_i\beta)^2 \right] + \lambda \sum_{j=1}^{p} |\beta_j|. \qquad (6)$$

The Lasso can be considered as an OLS estimator with a penalty $\lambda$ on the sum of the absolute coefficients. We obtain the standard OLS coefficients if the penalty term is zero and we have at least as many observations as covariates. For a positive penalty term, some coefficients are shrunken towards zero to satisfy the constraint. Thus, the Lasso serves as a variable selector because some variables have their coefficients set exactly to zero if the penalty is gradually increased. By increasing the penalty term to a sufficiently large number, one can obtain a path from a full model to an empty model with all coefficients besides the constant being zero. The idea of this procedure is to shrink those variables with little or no predictive power to zero and use either the remaining shrunken coefficients (Lasso), or the unshrunken coefficients from an OLS regression with the non-zero estimates (Post-Lasso) for prediction.

The controls $X_i$ entering equation (6) in our application is obtained in the following way. Starting from all second order interactions of the 328 base variables and 4th order polynomials for continuous variables, we drop those interactions that create empty cells or nearly empty cells containing less than 1% of the observations. We keep only one variable of variable groups that show absolute correlations above 0.99. Finally, we add dummies for states, school track, and each school in the sample. This gives a total of 10,066 variables to be considered in the selection process.

Dummies for state and school track are left unpenalized because institutional knowledge tells us that we expect substantial differences across states and school tracks, which should be accounted for. The "empty" model therefore contains already 19 variables and the Post-Lasso is used to find

---

[12] See for extensive treatments of the Lasso, e.g., Bühlmann and van de Geer (2011) and Hastie et al. (2015).

the predictors that should enter on top.[13] The penalty term that determines how many additional controls enter the models of the nuisance parameters is chosen via 10-fold cross-validation of Post-Lasso. This procedure aims to find the penalty term that minimizes the out-of-sample mean squared error (MSE) and is standard in nonparametric and machine learning estimation (see for a general review, Arlot & Celisse, 2010). The details are described in Appendix B

The outcome predictions are obtained by separate OLS Post-Lasso regressions in each treatment category. The predicted propensity score is obtained by separate logistic Post-Lasso regressions to account for the binary nature of the treatment indicators (Belloni, Chernozhukov, & Wei, 2013). Appendix E shows the obtained propensity scores and explains how we enforce common support.

As stated in the previous section, the estimator requires that predictions of the nuisance parameters converge at the rate $n^{-1/4}$. At this stage, we need to assume that the cross-validated Post-Lasso achieves this rate because the convergence rate of this particular estimator is not yet available. The assumption that this convergence rate is feasible builds on two theoretical results. First, Chetverikov, Liao and Chernozhukov (2017) show that cross-validated Lasso can reach $n^{-1/4}$ convergence assuming sparsity of the underlying model. The sparsity assumption means that the number of relevant variables $s$ is much smaller than the number observations $n$.[14] We are not aware of any theoretical or heuristic tests for the plausibility of sparsity assumptions. However, our sensitivity analysis regarding the penalty choice provides evidence that the sparsity assumption is not an issue in this application. Sparsity might be seen as conceptualization of the empirical practice to include only few controls compared to sample size. Such practices implicitly assume that the chosen variables are sufficient to provide a good approximation of the models of interest. Second,

---

[13] A sensitivity analysis without forcing these dummies into the model shows very similar results to this procedure.

[14] Formally, Chetverikov et al. (2017) show that cross-validated Lasso with Gaussian errors converges at the rate $(s \log p/n)^{1/2} \log^{7/8}(pn)$, where $p$ is the number of potential variables. This implies that $n^{-1/4}$ convergence requires $s^2 \log^2 p \log^{7/2}(pn)/n \to 0$. The sparsity requirements with cross-validated penalty are thus more strict compared to Lasso with data-driven penalty based on asymptotic arguments where $s^2 \log^2 p /n \to 0$ is needed for $n^{-1/4}$ convergence (Belloni et al., 2014b).

Belloni and Chernozhukov (2013) show that Post-Lasso converges at least as fast as Lasso under data-driven penalty terms based on asymptotic arguments. It seems therefore plausible to assume that a similar relation holds also for cross-validated Post-Lasso and cross-validated Lasso such that the required convergence of $n^{-1/4}$ is feasible in our implementation.

Finally, we need to cluster the standard errors at school level *s* because the sampling is school based (von Maurice, Sixt, & Blossfeld, 2011). The clustered standard errors are estimated as $\hat{\sigma}_{\mu,t}/\sqrt{n}$ where

$$\hat{\sigma}^2_{\mu,t} = \frac{1}{N} \sum_s \left( \sum_{i \in s} \frac{d_i^t(Y_i - \hat{\mu}_t(X_i))}{\hat{p}_t(X_i)} + \hat{\mu}_t(X_i) - \hat{\mu}_t \right)^2. \qquad (7)$$

The hat notation in equation (7) indicates estimated sample equivalents of the arguments in equation (3). Corresponding to, e.g., clustered standard errors for OLS, equation (7) sums first over all students *i* in the same school *s* to account for potential within-school correlations before summing over the schools.

## 4.4 Assessment of covariate balancing

Good practice in (multivalued) treatment effects applications based on propensity scores requires to assess the balancing of the covariates before and after adjusting for selection (Imbens & Wooldridge, 2009). These balancing checks exploit that the estimate of the mean potential outcome of treatment group *t* can be expressed as a weighted average of the observed treated or formally as $\hat{\mu}_t = Y_t w_t^p$, where $Y_t$ is a 1 x $N_t$ vector containing the $N_t$ observed outcomes in this treatment group and $w_t^p$ is an $N_t$ x 1 vector containing weights obtained from matching or weighting by the propensity score (see, e.g., Huber, Lechner, & Wunsch, 2013; Smith & Todd, 2005). For example, $w_t^{p'} = [w_{t,1}^p, ..., w_{t,N_t}^p]$ with $w_{t,i}^p = d_i^t/\hat{p}(x)$ for inverse probability weighting (Hirano, Imbens, & Ridder, 2003; Horvitz & Thompson, 1952). Balancing of the covariates between different treatment groups is then assessed based on the weighted covariates $X_t w_t^p$, where $X_t$ is a $p \times N_t$ matrix containing the $p$ covariates of the observations in treatment group $t$ (W. S. Lee, 2013).

So far, balancing tests are not conducted for estimators based on efficient scores. Though not naturally appearing in the estimation procedure, the underlying weights can be calculated as soon as a weighted representation of the predicted outcome is available as $\hat{\mu}_t(X_i) = Y_t w_t^Y$. The empirical version of equation (2) can then be rewritten as

$$\hat{\mu}_t = \frac{1}{n} \sum_{i=1}^{n} \left( \frac{d_i^t (Y_i - \hat{\mu}_t(X_i))}{\hat{p}_t(X_i)} + \hat{\mu}_t(X_i) \right) = \frac{1}{n} \sum_{i=1}^{n} \left( \frac{d_i^t Y_i}{\hat{p}_t(X_i)} - \frac{d_i^t Y_t w_t^Y}{\hat{p}_t(X_i)} + Y_t w_t^Y \right)$$

$$= Y_t w_t^p + Y_t w_t^Y - Y_t w_t^{pY} = Y_t (w_t^p + w_t^Y - w_t^{pY}) = Y_t w_t. \qquad (8)$$

The implementation via Post-Lasso allows us to calculate $w_t$ because the weights for predicting the outcome of unit $i$ are provided by the $N_t \times 1$ vector $w_{t,i}^Y = X_t (X_t' X_t)^{-1} X_i$ and sum to one (Abadie, Diamond, & Hainmueller, 2015).[15] To calculate the weight vector $w_t$, we need $w_t^Y = [w_{t,1}^Y, \ldots, w_{t,N}^Y]\mathbf{j}$, where $\mathbf{j}$ is a $N \times 1$ of ones, as well as $w_t^{pY} = [w_{t,1}^p w_{t,1}^Y, \ldots, w_{t,N}^p w_{t,N}^Y]\mathbf{j}$. The vector $w_t = w_t^p + w_t^Y - w_t^{pY}$ gives then the weight that each outcome in the treatment group receives in the estimation of the mean potential outcome. The weights $w_t$ can be used for balancing checks of the weighted covariates $X_t w_t$ or to detect extreme weights due to small propensity scores or extrapolation.[16] This is even more important in the high-dimensional setting of this paper because only few variables might be selected in the estimation but all confounders need to be balanced. The ability to validate that balancing works properly should thus be an integral component of the analysis. These checks are good empirical practice but they come at the cost of limiting the choice of estimators for $\mu_t(x)$ to methods with a known weighted representation.

---

[15] To the best knowledge of the author, there is currently no such weighted representation of the standard Lasso available. Thus, balancing checks for an estimator using standard Lasso predictions would not be possible. Feasible alternatives are post-Boosting (Luo & Spindler, 2016) or Random Forests (Breiman, 2001).

[16] The weighted representation works for all the estimators based on the efficient score for average treatment effects like the efficient influence function estimator (Cattaneo, 2010), augmented inverse probability weighting (Glynn & Quinn, 2009), or the DML approaches for binary treatments (Chernozhukov, Chetverikov, et al. 2017, 2018).

## 4.5 Sensitivity analysis regarding penalty choice

The paper of Farrell (2015) concludes by emphasizing the importance of the penalty parameter and the lack of knowledge about the proper choice. Thus, it is important to assess the credibility of the results by checking the sensitivity of the results to the penalty choice.

We propose one way to systematically address this issue. It shows the advantage of using cross-validated penalty terms instead of theoretical ones because they allow a data-driven assessment of sensitivity.[17] It is based on the one-standard-error rule (1SE) introduced by Breiman et al. (1984) in the context of cross-validation of classification and regression trees. The 1SE rule is motivated by the observation that the cross-validated MSE is rather similar around the penalty value that indicates the minimum cross-validated MSE.[18] Consequently, there is some degree of uncertainty about the MSE minimizing penalty and the model complexity might vary substantially over plausible values of the penalty term. Breiman et al. (1984) propose to estimate the standard error of the cross-validated MSEs and to take the penalty that is one standard error in the direction of a *smaller model*. Although the choice of one standard error is ad-hoc, the 1SE rule is widely applied and taught in machine learning textbooks (Hastie et al., 2009; Hastie, Tibshirani, & Wainwright, 2015). The underlying idea is that we want to opt for the less complex model under uncertainty about the optimal penalty term.

We propose to complement the 1SE rule by the 1SE+ rule that considers the *more complex model* within one standard error. These rules are particularly useful for estimators with multiple nuisance parameter that are obtained from different machine learners as in our case. The levels of penalties for least squares and logistic Lasso are not necessarily comparable. Therefore, running sensitivity checks by changing the penalty terms for all nuisance parameters by a fixed absolute or relative amount is problematic.[19] Instead, applying the alternative rules simultaneously to all nuisance estimators provides a data-driven way to investigate sensitivity of the estimates to the penalty choice. Additionally, this procedure naturally accounts for the possibility that the MSE minimizing penalty terms might be estimated with different precision. Those nuisance parameters with rather flat and imprecisely measure MSE curves vary more in this procedure than the precisely measured curves with a clear global minimum.

---

[17] A second advantage of using cross-validated instead of asymptotic penalties is the increased robustness to deviations from the theoretical setup used to derive the asymptotic penalties. Farrell (2015) notes in his simulation study that cross-validation provides excellent performance over a variety of sparse data generating processes, while the asymptotic choice is more sensitive.

[18] Appendix B provides a representative example and a formal description of the 1SE rule.

In section 5.3, we investigate the 1SE, 1SE+, and 2SE+ rules and compare their results to the cross-validated minimum.[20] Figure B.1 of Appendix B provides a representative example that shows how the different rules with similar magnitudes of cross-validated MSE show substantially different numbers of included variables. The investigation of results obtained using different penalties indicates whether or not the method produces stable results for a range of plausible penalty terms. In the ideal case, the estimates should be stable if model complexity is increased beyond the cross-validated minimum but the standard errors should get larger. This would indicate that the confounding is sufficiently controlled for at the cross-validated minimum and all additional variables just decrease efficiency.

Checking the sensitivity with regard to penalty term choice may be also informative about some other issues in the analysis. (i) The Post-Lasso estimator assumes sparsity. If going from the minimum penalty to more complex models changes the estimated effects substantially, this could be an indicator for a failure of sparsity in a specific application. (ii) The cross-validation optimizes the MSE of the treatment and outcome but not of the (unobserved) causal effect. Therefore, the procedure aims to minimize the MSE for the wrong estimand (see Frölich (2005) for a similar argument regarding non-parametric estimators as plug-ins for causal effects). Instability of the estimated effects around the minimum penalty could indicate that this concern is relevant.

---

[19] A 10% decrease in the penalty term could, e.g., lead to a large number of added variables in the outcome equation but only a few in the treatment equation.

[20] A 2SE rule leads in most cases to empty models in is not considered.

Although rules based on cross-validation standard errors are arbitrary and not based on theoretical considerations, they provide a systematic tool to assess sensitivity of causal estimates based on machine learning algorithms in practice.

# 5 Results

## 5.1 Variable selection and covariate balancing

Before discussing the estimated effects, we take a look at the variable selection in the machine learning step and the balancing performance of DML. Recall that the Lasso starts with 19 state and school track dummies. Panel A of Table 1 shows that less than ten variables are on average selected at the cross-validated minimum of the Post-Lasso in addition. The weights of equation (8) allow us to assess whether this rather small number of selected variables successfully balances the distribution of all ten thousand controls.

We follow Yang et al. (2016) and check this by calculating standardized differences (SD). These scale the mean difference between one treatment group and the other groups by the square root of the mean variances of all treatment groups and multiply this fraction by 100. We calculate SD for all intensity groups and look at the maximum absolute SD for each variable. Panel B of Table 1 provides summary statistics of the absolute SD for the binary and multiple treatment case before and after DML. The comparison before DML shows that some covariates are highly unbalanced with a maximum absolute SD larger than 30 and thus far above the 20 that are considered as being large by Rosenbaum and Rubin (1985).[21] However, most of the controls are decently balanced as documented by a mean absolute SD between three and four as well as by the fraction of variables with absolute SD above 10% being between 5% and 7%.

DML improves the balancing substantially and the few selected variables suffice to balance also the variables that are not selected. The maximum absolute SD is less than one third of the before value and far below 20 after DML adjustment. Furthermore, over 500 variables showed an SD above ten without adjustment. DML reduces this number to zero for the binary case and to less than ten in the multiple case. The effect estimates in the next section are thus not driven by large imbalances in the distribution of controls after adjusting for a small set of selected controls.[22] This insight would not be possible without the weighted representation and emphasizes the value of balancing checks especially in high-dimensional settings.

---

[21] Table A.3.1 of Appendix A.3 shows which baseline characteristics are the main drivers of imbalance.

*Table 1: Number of selected variables and covariate balancing*

|  | Binary treatment | | Multiple treatment | |
| --- | --- | --- | --- | --- |
|  | Before | After | Before | After |
| *Panel A: Number of selected variables* | | | | |
| (Mean) # of selected variables for treatment equation | - | 9 | - | 4.5 |
| Mean # of selected variables for outcomes equation | - | 7.7 | - | 4.6 |
| *Panel B: Balancing of all 10,066 covariates* | | | | |
| Maximum |SD| | 32.7 | 8.0 | 35.2 | 11.5 |
| Mean |SD| | 3.5 | 1.7 | 3.2 | 2.1 |
| Fraction of variables with |SD| > 10 in % | 5.2 | 0.0 | 6.5 | 0.1 |
| Fraction of variables with |SD| > 5 in % | 25.1 | 2.5 | 38.3 | 16.2 |

Note: Panel A shows numbers of additionally selected variables at the cross-validated minimum of Post-Lasso. The numbers for the propensity score of multiple treatments are average over all treatment states. The numbers of outcome predictions are averaged over all treatment states and outcomes. Panel B summarizes the absolute standardized differences (|SD|, Yang et al. 2016). The columns before are based on the unconditional differences. The after columns are calculated after DML adjustment using weights of equation (8) and averaged over all outcomes.

## 5.2 Effects of music on youth development

Table 2 shows the results for the comparison of musically active and inactive students in column one as well as comparisons between the different intensity categories in the remaining columns. Pairwise comparisons of intensities always compare the higher with the lower intensity in the respective pair. All outcome variables are standardized to have zero mean and variance one.

For *cognitive skills*, the first column reports highly significant increases of about 0.1 standard deviations (sd) for objectively measured science, math, vocabulary, and ICT skills for students practicing music at least one day per month. Only reading skills show no significant improvement.

---
[22] A visualization of the balancing improvement is provided in Appendix C.2.

This result is qualitatively in line with Cabane et al. (2016) and Hille and Schupp (2015). The latter show similar effect sizes for cognitive skill. With their standard errors being four times larger than those obtained in this study, they cannot report statistical significance, though. This might be mainly attributed to our substantially larger sample size.

*Table 2: Main results for binary and dose-response treatment effects*

|  | Binary | Dose-response | | | | | |
|---|---|---|---|---|---|---|---|
|  | Any - No | Low - No | Med - No | High - No | Med - Low | High - Low | High - Med |
|  | (1) | (2) | (3) | (4) | (5) | (6) | (7) |
| *Cognitive skills (standardized)* | | | | | | | |
| Science | 0.11*** | 0.04 | 0.14*** | 0.17*** | 0.10*** | 0.13*** | 0.03 |
|  | (0.02) | (0.03) | (0.03) | (0.04) | (0.04) | (0.04) | (0.04) |
| Math | 0.08*** | 0.05 | 0.12*** | 0.10*** | 0.07** | 0.05 | -0.02 |
|  | (0.02) | (0.03) | (0.02) | (0.04) | (0.04) | (0.04) | (0.04) |
| Vocabulary | 0.11*** | 0.02 | 0.16*** | 0.18*** | 0.14*** | 0.16*** | 0.02 |
|  | (0.02) | (0.03) | (0.03) | (0.03) | (0.03) | (0.04) | (0.04) |
| Reading | -0.03 | 0.01 | -0.04 | -0.01 | -0.06 | -0.02 | 0.03 |
|  | (0.02) | (0.04) | (0.03) | (0.04) | (0.04) | (0.05) | (0.04) |
| ICT | 0.12*** | 0.06* | 0.15*** | 0.18*** | 0.09** | 0.11** | 0.03 |
|  | (0.02) | (0.03) | (0.03) | (0.04) | (0.04) | (0.04) | (0.04) |
| *School performance (standardized)* | | | | | | | |
| German grade | 0.12*** | 0.11*** | 0.13*** | 0.16*** | 0.03 | 0.05 | 0.03 |
|  | (0.03) | (0.04) | (0.03) | (0.04) | (0.04) | (0.05) | (0.05) |
| Math grade | 0.05* | 0.04 | 0.08** | 0.04 | 0.04 | -0.003 | -0.04 |
|  | (0.03) | (0.04) | (0.04) | (0.05) | (0.04) | (0.05) | (0.05) |
| Average grade German & math | 0.09*** | 0.09** | 0.13*** | 0.10** | 0.03 | 0.01 | -0.03 |
|  | (0.03) | (0.04) | (0.04) | (0.04) | (0.04) | (0.05) | (0.05) |
| *Big Five (standardized)* | | | | | | | |
| Extraversion | 0.03 | 0.001 | 0.04 | 0.07 | 0.04 | 0.07 | 0.02 |
|  | (0.03) | (0.04) | (0.04) | (0.05) | (0.04) | (0.05) | (0.05) |
| Agreeableness | 0.11*** | 0.11*** | 0.10*** | 0.11*** | -0.005 | 0.006 | 0.01 |
|  | (0.03) | (0.04) | (0.04) | (0.04) | (0.04) | (0.05) | (0.05) |
| Conscientiousness | -0.04 | -0.04 | -0.06* | -0.007 | -0.02 | 0.04 | 0.05 |
|  | (0.03) | (0.04) | (0.03) | (0.04) | (0.04) | (0.05) | (0.05) |
| Neuroticism | 0.001 | 0.05 | -0.02 | -0.04 | -0.08* | -0.10* | -0.02 |
|  | (0.03) | (0.04) | (0.04) | (0.04) | (0.04) | (0.05) | (0.05) |
| Openness | 0.31*** | 0.13*** | 0.33*** | 0.50*** | 0.20*** | 0.37*** | 0.18*** |
|  | (0.03) | (0.04) | (0.03) | (0.04) | (0.04) | (0.05) | (0.05) |

Note: This table shows the estimated effects comparing different intensities of musical practice. All outcome variables are standardized to mean zero and variance one. Higher grades are better. The results are obtained by applying the Farrell (2015) estimator using Post-Lasso with penalty chosen at the minimum of 10-fold cross-validated MSE. State and school track dummies enter the selection unpenalized. Standard errors in parentheses are clustered at the school level. *, **, *** indicate statistical significance at the 10%, 5%, 1% level, respectively.

The comparison of different intensities shows a clear pattern. The improvements are mainly driven by students with medium and high intensities. For example, science skills improve by a highly significant 0.14 sd for medium intensity versus inactive students and 0.17 sd for high intensity versus inactive students. In contrast, the increase of 0.04 sd for low intensity versus inactive is not significant. Similar patterns are also observed for math, vocabulary, and ICT skills. Columns four and five document that the differences of the medium or high versus low intensity are also significant at least at the 5% level. The only exception is high versus low intensity for math skills. Column seven shows no further significant improvements for high intensity versus medium intensity practice.

The results on cognitive skills suggest that the cognitive benefits materialize only for serious practice and not for only occasional music making. This is in contrast to the finding for *school performance* in the panel below. In line with previous studies, column one shows significant improvements for German and math grades for musically active students, while the improvements of German are more pronounced with a highly significant 0.12 sd compared to a marginally significant 0.05 for math. Unlike in the case of cognitive skills, column two shows that even a low intensity of music results in significantly improved German grades. The comparisons between low, medium and high intensities in columns five to seven show no additional significant difference. One potential explanation of this pattern is that the mere signal of playing music is already rewarded by teachers (see for similar results and discussion Hille & Schupp, 2015). A potential explanation for different sizes of the effects is that low intensity students participate in school-based musical activities like voluntary school choirs and German teachers are more receptive to this signal than math teachers. This is plausible for two reasons. First, the subject of German is closer related to arts than math and German teachers potentially care more about artistic activities of their students than their math colleagues. Second, even if German and math teachers care to the same degree, German grades are more subjective compared to the relatively objectively measurable performance in math.

The results regarding the *Big Five* suggest that playing music has a significant effect on two personality traits. The measure of agreeableness increases by 0.11 for musically active students. This effect does not differ for different intensity categories. However, the large increase in openness by 0.31 sd for music versus no music can be split into differential increases with intensity. Openness is the only variable that shows a significant and large difference between high and medium intensity with 0.18 sd in column seven. The finding that openness is by far most affected by musical practice is in line with the results in Cabane et al. (2016) and Hille and Schupp (2015). Similar to their data, openness in the NEPS data is partly assessed by asking about artistic interests (Rammstedt & John, 2007; Wohlkinger, Ditton, von Maurice, Haugwitz, & Blossfeld, 2011). This could explain the observed effect sizes at least partly.

## 5.3 Sensitivity analyses

The results of DML methods might be sensitive to the *tuning parameter choice* in the machine learning part. However, the literature lacks guidance on how to assess the sensitivity to penalty choices. Section 4.5 proposes alternative rules for a systematic investigation of the sensitivity of the results to smaller or larger models. Appendix D.1 discusses in detail how important parameters of the analysis like number of selected variables, implied weights and covariate balancing vary for different penalty term choices and how these differences affect the results. The main conclusion is that increasing the model complexity beyond the baseline MSE minimizing penalty leads to only marginal changes of the results. This is surprising as the alternative penalty choice rules select up to six times more additional control variables than the baseline penalties. However, we observe that the rather sparse specifications of the baseline with not more than ten additional variables seem already sufficient to control for the selection into playing music. The estimated effects obtained from more complex models vary only within one standard error of the baseline results and the qualitative conclusions are the same. One exception is the effect on math grades for the binary treatment case. This effect is significant at 10% in the baseline but not anymore when including more control variables. We find in general that adding more variables than in the baseline model increases covariate balancing of all potential confounders only marginally but decreases efficiency mildly due to more extreme weights.

It would be tempting to run an additional sensitivity analysis that compares the baseline results with results from only inverse probability weighting (IPW) and only regression adjustment (RA). However, these single equation approaches do not yield uniformly valid statistical inference when we apply machine learning as discussed in section 2. Thus, we would not know how to calculate standard errors for these separate approaches. Instead, we use the "anatomy" of DML weights in equation (8) to investigate how the IPW weights ($w_t^p$) and RA weights ($w_t^Y$) relate to the DML weights ($w_t$). The results in Appendix D.2 show that DML weights are highly correlated with the IPW weights. Further, we observe that the rather poor balancing performance of the RA weights is offset by the adjustment weight ($w_t^{pY}$). The DML weights lead to slightly better balancing compared to the balancing obtained from IPW weights only. Thus, in addition to valid statistical inference, the DML approach also improves balancing of the covariate distribution compared to the single equation approaches.

Appendix D.3 discusses further sensitivity analyses in detail. Including the potentially endogenous control variables to the set of available controls produces very similar results compared to the baseline. Also restricting the comparison to music versus sports instead of music versus any kind of extracurricular activities does not alter the qualitative findings. However, using the full sample and comparing music versus all non-musicians produces several more significant positive effects that might be explained by the failure to control for selection into any extracurricular activity. This emphasizes the importance of the approach advocated in Cabane et al. (2016).

Sensitivity checks regarding the enforcement of common support and the fixed inclusion of state and school track dummies show that these choices do not matter for the results either. Surprisingly, considering only the main effects instead of the large set of interactions and polynomials also produces results that are very similar to the baseline with over 10,000 available potential variables.

Finally, Appendix D.3.6 investigates whether the results are robust to the use of an alternative machine learning algorithm to estimate the nuisance parameters. Especially the untestable sparsity assumptions required for the Lasso estimators might be considered as critical. Thus, we apply the Random Forest estimator (Breiman, 2001) as a flexible alternative that requires no sparsity assumptions. The obtained results are very similar to the main results.

# 6   Conclusion

This study investigates the effect of playing music on the cognitive and non-cognitive skills of 9$^{th}$ grade students in Germany. The results are in line with previous studies showing significantly positive effects of musical practice *per se*. Going beyond the mere comparison of musicians and non-musicians, the study assesses the effects of different intensity levels of practice. It is shown that standardized and objectively measured cognitive skills require at least a medium level of practice to show notable benefits. However, substantial improvements in teacher assessed German grades are already observed for low intensity practice. This is in line with similar observations in Hille and Schupp (2015) who argue that playing music might affect school grades also through a positive signal to teachers. Regarding non-cognitive skills, we document significant improvements in openness that are increasing with intensity level. However, the openness indicator creates a mechanical relation to music by including artistic interests. Therefore, it remains an open question whether similar effects would be found for different indicators of openness. Overall, we find no evidence that a high intensity of making music could be harmful by crowding-out other important activities.

The estimation of the effects is implemented via recent DML estimators that allow a flexible and transparent way to obtain causal estimates in observational studies. One concern regarding these methods is that they might depend heavily on the specific parameter choice. This paper proposes a systematic way to address these concerns based on cross-validation of Post-Lasso. The procedure provides important insights and is of general use for DML with all machine learners that are tuned via cross-validation. The sensitivity analysis finds stable results for a range of plausible penalty terms. Maybe surprisingly, very small models that include only about 10 variables suffice to obtain stable effects and moving to substantially richer model specifications leads only to mildly decreased efficiency. Another interesting point is that considering only the baseline characteristics as potential control variables instead of the more flexible set with interactions and polynomials gives nearly identical results. This indicates that increasing the flexibility and the computational burden is not necessary in this application. However, this could be different for other applications.

The paper derives a weighted representation of the DML estimator that has proven to be useful to incorporate standard empirical practices regarding covariate balancing checks in applications. It generalizes to all machine learners where the predictions can be written as weighted averages of the outcomes. Thus, practitioners might face a trade-off between the possibility of checking covariate balancing and the use of sophisticated machine learners with unknown weighted representation.

On the methodological side, further research is required to investigate how different choices made throughout the paper are sensible. The goal should be to find good practices for DML in empirical applications. These are needed for all details of the implementation, especially regarding different choices of predictors and penalty terms, the dimension of the covariate matrix (order of interactions and polynomials), and common support enforcement. Further, the relevance of the idea of cross-fitting proposed by Chernozhukov, Chetverikov, et al. (2018, 2017) should be investigated.[23] Regarding identification, the available data about parental tastes seem to be crucial. However, future investigations should add better measures of early ability to check whether those factors are driving the mostly positive results of extracurricular activities in the literature.

---

[23] The data for this project are only available via remote access and the computationally more expensive cross-fitting is therefore not pursued in this project.

*Neuroscience*, *7*, 1–9.

Bertrand, M., Crépon, B., Marguerie, A., & Premand, P. (2017). *Contemporaneous and post-program impacts of a public works program: Evidence from Côte d'Ivoire* (Working Paper). Washington D.C.

Bischoff, S. (2011). *Deutsche Musikvereinigungen im demographischen Wandel - zwischen Tradition und Moderne* (Schriftenreihe der Bundesvereinigung Deutscher Orchesterverbände e.V. No. 5).

Blossfeld, H.-P., Roßbach, H.-G., & von Maurice, J. (2011). Education as a lifelong process – The German National Educational Panel Study (NEPS). *[Special Issue] Zeitschrift Für Erziehungswissenschaft*, *14*.

Breiman, L. (2001). Random forests. *Machine Learning*, *45*(1), 5–32.

Breiman, L., Friedman, J., Stone, C. J., & Olshen, R. A. (1984). *Classification and regression trees*. CRC press.

Bühlmann, P., & van de Geer, S. (2011). *Statistics for high-dimensional data: Methods, theory and applications*. Springer Science & Business Media.

Cabane, C., Hille, A., & Lechner, M. (2016). Mozart or Pelé? The effects of adolescents' participation in music and sports. *Labour Economics*, *41*, 90–103.

Cattaneo, M. D. (2010). Efficient semiparametric estimation of multi-valued treatment effects under ignorability. *Journal of Econometrics*, *155*(2), 138–154.

Chernozhukov, V., Chetverikov, D., Demirer, M., Duflo, E., Hansen, C., & Newey, W. (2017). Double/Debiased/Neyman machine learning of treatment effects. *American Economic Review Papers and Proceedings*, *107*(5), 261–265.

Chernozhukov, V., Chetverikov, D., Demirer, M., Duflo, E., Hansen, C., Newey, W., & Robins, J. (2018). Double/Debiased machine learning for treatment and structural parameters. *The Econometrics Journal*, *21*(1), C1–C68.

Chernozhukov, V., Goldman, M., Semenova, V., & Taddy, M. (2017). *Orthogonal Machine Learning for Demand Estimation: High Dimensional Causal Inference in Dynamic Panels*. Retrieved from http://arxiv.org/abs/1712.09988

Chernozhukov, V., Hansen, C., & Spindler, M. (2016). *High-Dimensional Metrics in R*. Retrieved from http://arxiv.org/abs/1603.01700

Chernozhukov, V., Newey, W., & Robins, J. (2018). *Double/De-Biased Machine Learning Using Regularized Riesz Representers*. Retrieved from http://arxiv.org/abs/1802.08667

Chetverikov, D., Liao, Z., & Chernozhukov, V. (2017). *On cross-validated Lasso*. https://doi.org/10.1920/wp.cem.2016.4716

Corrigall, K. A., Schellenberg, E. G., & Misura, N. M. (2013). Music training, cognition, and
29

# *Appendices*

*The following Appendices are not meant for publication.*

## Appendix A: Data

### A.1 Data preparation

This Appendix gives more details about the sample selection and data preparation steps. It shows two major reasons for dropping more than half of the sample. The first and most severe one is non-response of parents to the telephone interview. As detailed parental background information is of utter importance for the identification via the CIA, observations without parental information are discarded. The second major reason for omitting observations is non-response to the measure of musical practice. This is mainly driven by special needs schools where this question was not asked. Such schools teach adolescents with learning disabilities. The analysis is therefore restricted to children without learning disabilities.

*Table A.1.1: Sample selection*

| Step | Remaining # of observations |
|---|---|
| All students in the NEPS database | 15,577 |
| Merging with parents (missing parental interview) | 8,786 |
| Missing information about musical practice (mostly driven by special needs schools where the question was not asked) | 7,784 |
| Missing grades | 7,527 |
| Missing Big Five | 6,927 |
| Missing cognitive skills | 6,898 |
| No extracurricular activities | 5,943 |

Table A.1.2 on the next page describes the generation of all potential control variables that we consider from the NEPS data. Mainly categorical variables are coded as binary indicators for each category. Further, continuous variables such as wealth are additionally coded as categorical variables with binary indicators for each category.

*Table A.1.2: Coding of all considered control variables*

| | Coding of variables | # of generated variables | Potentially endogenous |
|---|---|---|---|
| **Student characteristics** | | | |
| Female | binary | 1 | |
| Recommendation secondary school | 4 categories | 4 | |
| Held back a year / repeated grade | binary | 1 | x |
| Skipped grade | binary | 1 | x |
| **Leisure preferences parents** | | | |
| Parent's quantity reading on leisure days (hrs/day) | continuous | 1 | |
| Number of books in HH | 6 categories (0-10 books to more than 500 books) | 6 | |
| Participation in high culture: museum, art exhibition | | 5 | |
| Participation in high culture: cinema | 5 categories (never last year - more than 5 times) | 5 | |
| Participation in high culture: opera | | 5 | x |
| Participation in high culture: theatre | | 5 | |
| Participation in high culture: rock concert | | 5 | x |
| **Work characteristics parents** | | | |
| Parent's quantity reading on work days (hrs/day) | continuous | 1 | |
| Mother employment status | 4 categories (unemployed, side-job, part-/ full-time) | 4 | |
| Father employment status | | 4 | |
| Mother occupation | 12 categories | 12 | |
| Father occupation | | 12 | |
| **Economic condition household** | | | |
| Assessment economic HH situation | 5 categories (very poor - very god) | 5 | |
| Assets in the HH: savings book/checking account | | | |
| Assets in the HH: building loan contract | | | |
| Assets in the HH: life insurance policy | 8 binary variables respectively + mutually exclusive groups describing all observed combinations of asset types | 166 | |
| Assets in the HH: fixed-interest securities | | | |
| Assets in the HH: stocks, funds, bonds | | | |
| Assets in the HH: business assets | | | |
| Assets in the HH: owner-occupied real estate property | | | |
| Assets in the HH: other real estate property | | | |
| Household assets not including debt in € | 1 continuous variable + 6 categorical (0 - > 500,000 €) | 7 | |
| HH debt in € | 1 continuous variable + 5 categorical (0 - > 200,000 €) | 6 | |
| HH net wealth in € | continuous difference of two variables above | 1 | |
| HH receives transfer payments | binary | 1 | |
| **Parenting attitudes** | | | |
| Interference in partner selection | | 4 | |
| Men & women same right to decide on family income | 4 categorical (completely disagree - strongly agree) | 4 | |
| Vocational training more important for boys than for girls | | 4 | |
| Wish about final degree of child | 4 categories (don't care - university) | 4 | x |
| Pocket money in € per month | continuous | 1 | |

Table continues on next page >

*Table A.1.2 continued*

|  | Coding of variables | # of generated variables | Potentially endogenous |
|---|---|---|---|
| Parents idealistic aspiration apprenticeship | 4 categories | 4 | x |
| Parents idealistic aspiration school graduation | 4 categories | 4 | x |
| Parents importance of good grades | 6 categories | 6 | x |
| Parents importance of professional success | 6 categories | 6 | x |
| In your house is there … | | | |
| … a desk to study | binary | 1 | |
| … a room just for student | binary | 1 | |
| … learning software | binary | 1 | |
| … classic literature | binary | 1 | |
| … books with poems | binary | 1 | |
| … works of art | binary | 1 | |
| … books useful for homework | binary | 1 | |
| … a dictionary | binary | 1 | |
| Household demographics | | | |
| Age father | 1 continuous + 7 age categories of 5 years + 1 category if missing | 9 | |
| Age mother | | 9 | |
| Highest education mother | 10 categories | 10 | |
| Highest education father | 10 categories | 10 | |
| Marital status | 6 categories | 6 | |
| Migration background | binary | 1 | |
| Household size | continuous | 1 | |
| People under the age of 14 in HH | continuous | 1 | |
| Household composition: | | | |
| biological mother, adoptive mother, foster mother | binary | 1 | |
| stepmother or father's girlfriend | binary | 1 | |
| Biological father, adoptive father, foster father | binary | 1 | |
| Stepfather or mother's boyfriend | binary | 1 | |
| Siblings and/or stepsiblings | binary | 1 | |
| Grandmother and/or grandfather | binary | 1 | |
| Social circle of parents includes … | | | |
| … nurse or male nurse | binary | 1 | x |
| … engineer | binary | 1 | x |
| … warehouse / transport worker | binary | 1 | x |
| … social worker | binary | 1 | x |
| … sales clerk | binary | 1 | x |
| … police officer | binary | 1 | x |
| … doctor | binary | 1 | x |
| … banker | binary | 1 | x |
| … car mechanic | binary | 1 | x |
| … legal practitioner | binary | 1 | x |
| … optician | binary | 1 | x |
| … translator | binary | 1 | x |
| … teacher | binary | 1 | x |
| Regional information | | | |
| Population density | 7 categories | 7 | |
| Music school in district | binary | 1 | |
| Number of variables | | 377 | 49 |

*Note: HH means household, hrs means hours.*

## A.2 Extracurricular activities

Table A.2.1 shows characteristics of extracurricular activities by intensity category for a better understanding of what playing music means and how musically inactive students spend their time. The first row shows that the high intensity group reports on average twice the number of days of musical practice compared to medium intensity. The second row displays the fraction of students in each group that reports taking lessons at a music school. Only 21% of the low intensity group report taking lessons at a music school, while 52% and 68% in the medium and high intensity groups, respectively. This observation is in line with the argument that low intensity must not mean serious practice, assuming taking formal music lessons at a music school is a good indicator for serious practice. These are the only available variables that can be used to characterize the nature of musical engagement. However, it is plausible to assume that the majority of the remaining active students participates in music clubs (*"Musikverein"*) in which about half a million children and adolescents participate in Germany (Bischoff, 2011). Most of them also offer music lessons but would not be counted as music schools.

After a description of musical activity, the remainder of Table A.2.1 shows other extracurricular activities that are assessed in the questionnaire: sports, voluntary relief organizations, religious youth groups, fan clubs, culture clubs, and political associations. Music seems not to crowd out any of the other extracurricular activities besides activities in fan clubs, which are substantially less popular among musicians. As a consequence, musicians engage in on average more extracurricular activities with 2.5 compared to 1 for non-musicians. This difference is mainly driven by 26% of the non-musicians who do not participate in any of these extracurricular activities. The question is how do non-musicians spend their time? One part of the answer is given in the last three rows of Table A.2.1. Non-musicians are twice as likely to report that they play at least two hours online-role PC-games, skill or strategy PC-games, or other PC-games on normal school days.

*Table A.2.1: Description of extracurricular activities of students by different intensity levels*

|  | No music | Music | | | |
|---|---|---|---|---|---|
|  |  | Any | Low | Medium | High |
| Days/month musical practice | 0.00 | 14.90 | 3.56 | 14.48 | 28.35 |
| Taking music lessons at music school | 0.03 | 0.47 | 0.21 | 0.52 | 0.68 |
| Other extracurricular activities: |  |  |  |  |  |
|     Sports club | 0.65 | 0.68 | 0.69 | 0.70 | 0.67 |
|     Voluntary relief organizations | 0.11 | 0.12 | 0.11 | 0.12 | 0.12 |
|     Religious youth groups | 0.13 | 0.30 | 0.24 | 0.33 | 0.34 |
|     Fan clubs | 0.12 | 0.09 | 0.10 | 0.08 | 0.09 |
|     Culture clubs | 0.04 | 0.26 | 0.14 | 0.31 | 0.33 |
|     Political associations | 0.02 | 0.03 | 0.02 | 0.03 | 0.03 |
| Number of extracurricular activities | 1.06 | 2.48 | 2.29 | 2.57 | 2.57 |
| No extracurricular activity | 0.26 | 0.00 | 0.00 | 0.00 | 0.00 |
| At least 2 hours at normal school day playing … |  |  |  |  |  |
|     … online-role PC-games | 0.10 | 0.05 | 0.06 | 0.05 | 0.06 |
|     … skill or strategy PC-games | 0.08 | 0.05 | 0.04 | 0.05 | 0.05 |
|     … other PC-games | 0.21 | 0.12 | 0.13 | 0.10 | 0.13 |
| Number of observations | 3,582 | 3,316 | 1,027 | 1,369 | 920 |

Note: Table shows mean values of student characteristics by different intensity levels. All variables besides days/month musical practice and number of extracurricular activities are binary.

## A.3 Identification

This Appendix provides detailed arguments for the plausibility of the identification strategy that is outlined in section 4.1.

The remaining selection that needs to be controlled for after conditioning on active students concerns the decision to engage in music and not in something else as well as the intensity decision conditional on making music. Following the extensive discussions in Hille and Schupp (2015) and Cabane et al. (2016), the biggest driving factor is most likely parents who decide together with the child which activities to start. Especially *parental tastes* and *parenting attitudes* could influence activity choice but also many other outcomes of children. The parental survey of the NEPS provides a battery of different measures that could be useful to control for these parental characteristics. The information about leisure activities (museum, cinema, opera, etc.) and information about home possessions can account for parental tastes. In particular, the information about home possessions contains important measures of the cultural interests of parents. It seems plausible to assume that the availability of artwork, classic literature, or books with poems in the household measure the revealed cultural preferences of parents. These variables are not available in previous studies but may be crucial confounders, as shown below in Table A.3.1. Further, information about pocket money, the availability of learning materials or parental aspirations for the child can be used as measures of parenting attitudes.[24]

Correlated with the previous "soft" factors are the *economic conditions* of a household. Music lessons are more prevalent in families with higher socio-economic status in part because they can afford such extracurricular activities. Therefore, it is important that we observe all the "hard" information on wealth, debt, education, etc. as well as *parents work characteristics* and other *household demographics*. The *social circle* of the parents might also impact students' experience and maybe their decision for or against particular activities. Also in that dimension, the NEPS provides detailed information by asking about the occupations that are in the social circle of parents.

---
[24] The full list of variables in each category in italic is given in Table A.1.2.

Not only parental but also *student characteristics* could be of crucial importance. Especially early ability should be controlled for. However, the NEPS provides no direct measure of early ability. Thus, we follow previous studies and include recommendation for upper secondary school by teachers as proxy for early ability. Additionally, gender is included in the set of controls as girls are much more likely to play music as we see below.

Finally, *regional* and school characteristics are taken into account. The 16 German states have high independence in setting up their schooling systems. Therefore, state dummies and dummies for the secondary school tracks students are enrolled in (basic, intermediate, or academic track) are considered as control variables. Additionally, we add dummies for each school in the sample to account for unobserved peculiarities at the school level.[25]

While the above mentioned factors are all plausible confounders, it is interesting to investigate which are the main drivers of selection. We do this by calculating standardized differences (SD) like in section 5.1.

Table A.3.1 reports gender as the largest driving factor into music. Only 38% of the non-musicians are female compared to 60% of those students doing any music. Interestingly the female share is not increasing with intensity. The shares in the low and medium intensity are 61% and 63% and thus significantly higher compared to 55% in the high intensity group.

---

[25] The NEPS conducted a variety of school characteristics in an extensive survey. However, non-response to this survey would decrease the number of schools that we observe substantially. Thus, we stick to the inclusion of school dummies that should capture unobserved characteristics like musical profiles of the school, number of music teachers, etc if necessary.

*Table A.3.1: Mean comparison of control variables with standardized differences above 15*

|  | Largest \|SD\| | No music | Music | | | |
|---|---|---|---|---|---|---|
|  |  |  | Any | Low | Medium | High |
|  | (1) | (2) | (3) | (4) | (5) | (6) |
| Female | 32.3 | 0.38 | 0.60 | 0.61 | 0.63 | 0.55 |
| Classic literature in HH | 31.1 | 0.39 | 0.60 | 0.54 | 0.60 | 0.67 |
| Books with poems in HH | 25.2 | 0.59 | 0.76 | 0.72 | 0.77 | 0.79 |
| Academic track | 25.0 | 0.39 | 0.56 | 0.49 | 0.58 | 0.61 |
| Recommendation for academic track | 20.7 | 0.39 | 0.54 | 0.47 | 0.55 | 0.59 |
| More than 500 books in HH | 19.8 | 0.12 | 0.22 | 0.18 | 0.23 | 0.25 |
| Mother university degree | 18.6 | 0.08 | 0.16 | 0.12 | 0.18 | 0.18 |
| Father university degree | 18.3 | 0.10 | 0.19 | 0.14 | 0.21 | 0.21 |
| Never went to museum last year | 16.9 | 0.34 | 0.23 | 0.28 | 0.21 | 0.20 |
| 26 to 100 books in HH | 16.0 | 0.31 | 0.22 | 0.27 | 0.20 | 0.18 |
| Number of observations |  | 2,627 | 3,316 | 1,027 | 1,369 | 920 |

Note: |SD| means absolute standardized difference. These are shown in column (1) as mean difference between one group and the rest divided by the square root of the sum of the variances in both groups times 100. Variables are ordered by the maximum absolute SD observed for the four intensity groups. HH means household. Columns (2) – (6) show means of binary variables by intensity groups. Column (2) considers only extracurricularly active students and omits completely inactive students. Only variables not considered as potentially endogenous are included.

The other major drivers into music are parental cultural preferences, parental education and ability of children. Musically active children are more likely to live in households with classic literature, books with poems and a total number of books above 500. In contrast, non-musicians are more likely to live with parents that never went to a museum last year and have only fewer books at home. These cultural parental preferences are of course highly correlated with parental education, which also differs substantially. Parents of musically active students are twice as likely to hold a university degree compared to parents of non-musicians. Finally, students being in the academic track and having received such a recommendation in their early years are substantially more musically active. The observed patterns besides gender are all more pronounced for higher intensities of musical practice.

Note that all variables indicating economic conditions show SD below 10. They all go in the direction suggested by parental education, namely that children with wealthier parents are more likely to play music. However, the larger drivers seem to be parental preferences. This emphasizes the importance of measuring and including these preferences in the analysis.

# Appendix B: Cross-validation

The penalty term $\lambda$ is the crucial tuning parameter of the (Post-)Lasso as it determines the number of selected controls. We choose $\lambda$ via 10-fold cross-validation. This means, (i) the sample is randomly split into ten folds ($k = 1, \ldots, 10$) of similar size, (ii) the Lasso coefficient path is obtained over a grid of penalty terms, $\lambda \in \{\lambda_1, \ldots, \lambda_m, \ldots, \lambda_{100}\}$, in nine of these parts leaving out fold $k$,[26] (iii) standard OLS or logit coefficients are calculated in this sample using only the controls with non-zero coefficients at each grid point, (iv) these coefficients are used to predict values in the left out subsample at each grid point, $\hat{y}_{\lambda_m}^{-k}$, and (v) the cross-validated MSE of these predictions is calculated as $CV_k(\lambda_m) = n_k^{-1} \sum_i (y_i - \hat{y}_{i,\lambda_m}^{-k})$, where $n_k$ is the number of observations in the $k$-th fold. Steps (ii) to (v) are repeated ten times such that each subsample is used once as the left out sample. This provides ten series of MSE's over the whole penalty grid. Finally, we take the mean over the ten folds, $CV(\lambda_m) = 10^{-1} \sum_k CV_k(\lambda_m)$. For the baseline results we choose the penalty term $\lambda_{min}$ that minimizes the average MSE, $\lambda_{min} = \underset{\lambda \in \{\lambda_1, \ldots \lambda_{100}\}}{\mathrm{argmin}}\ CV(\lambda)$. $\lambda_{min}$ is then used to estimate the model in the full sample and to get the predictions that are plugged into the efficient score.

To obtain the one-standard-error rule (1SE), Breiman et al. (1984) propose to estimate the standard error of the cross-validated MSEs, $SE(\lambda_m) = \sqrt{\mathrm{var}(CV_1(\lambda_m), \ldots, CV_{10}(\lambda_m))/10}$. They start now from $\lambda_{min}$ and go along the penalty grid into the direction of a smaller model to find the first penalty $\lambda_{1SE}$ with $CV(\lambda_{1SE}) \leq CV(\lambda_{min}) + SE(\lambda_{min})$. 1SE+ or 2SE+ rules proceed the same but start the search into the direction of more complex models.

Figure B.1 shows a representative example of the cross-validated mean squared error and indicates how big the differences in the number of selected variables for reasonable penalty choices

---

[26] The grid of the 100 candidate penalty terms is chosen such that at most 500 variables enter the model to increase computational speed. At this point the model is always already too complex and shows bad out-of-sample performance indicating severe overfitting.

along the rather flat region around the cross-validated minimum might be. For example, the 1SE rule selects only three additional variables but the 1SE+ rule 17.

*Figure B.1: Representative example of cross-validation*

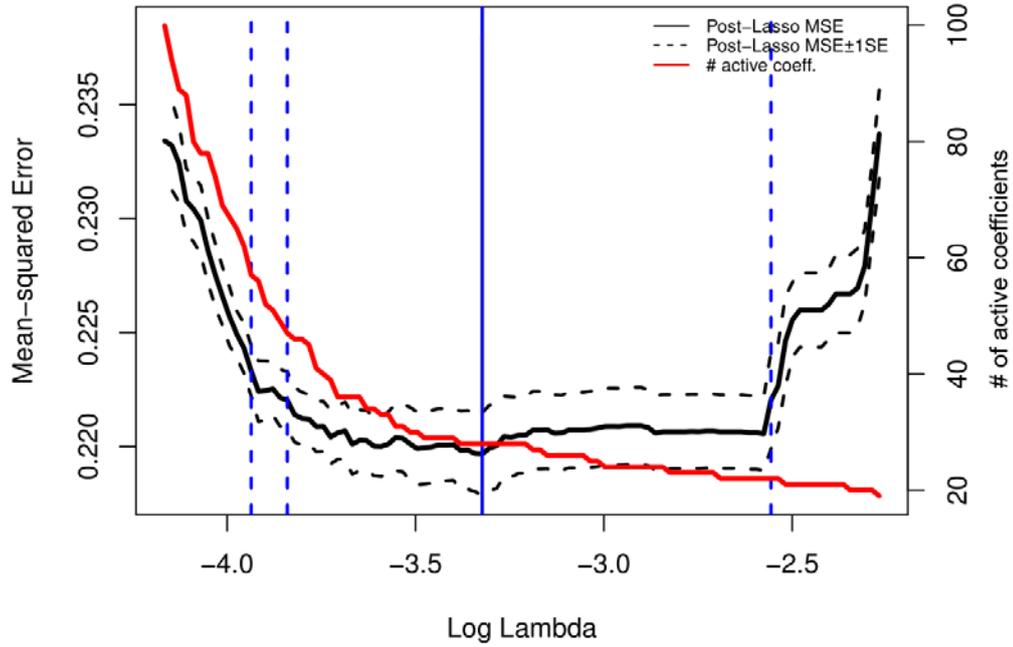

Notes: Cross-validation of propensity score for the binary treatment. It shows the mean-squared error along the penalty grid (black line), the number of variables included (red line) as well as the position of the 2SE+, 1SE+, minimum and 1SE penalty term choices (blue lines from left to right). Recall that the empty model contains already 19 variables.

# Appendix C: Assessment of balancing in high-dimensional settings

The Appendix in standard applications contains often an extended version of Table A.3.1 that shows standardized differences for all variables in the propensity score before and after adjustment. Such tables cover sometimes several pages but have finite length. The length of such tables would explode in the high-dimensional case because we do not just want to assess balance for the selected variables but especially want to check whether the potentially small set of selected variables also balances those variables that were not selected. The figures below visualize the improvement of balancing for all 10,066 variables. The black area shows the absolute standardized difference *before* adjustment ranked from the highest to the lowest values. The grey area shows the corresponding values *after* adjustment. As a representative example, Figures C.2.1 and C.2.2 shows balancing in the binary and multiple treatment case for science skills. We observe that only a minority of the controls show large imbalances before adjusting for selection. This observation is in favour of the sparsity assumption needed for the machine learning part to converge fast enough. Also the observation that the balancing of all variables is substantially improved by considering only a small subset can be interpreted as evidence for sparsity because otherwise controlling for only less than 30 variables could result in large differences of the variables that were not included.

*Figure C.2.1: Balancing before and after adjustment for selection – binary treatment*

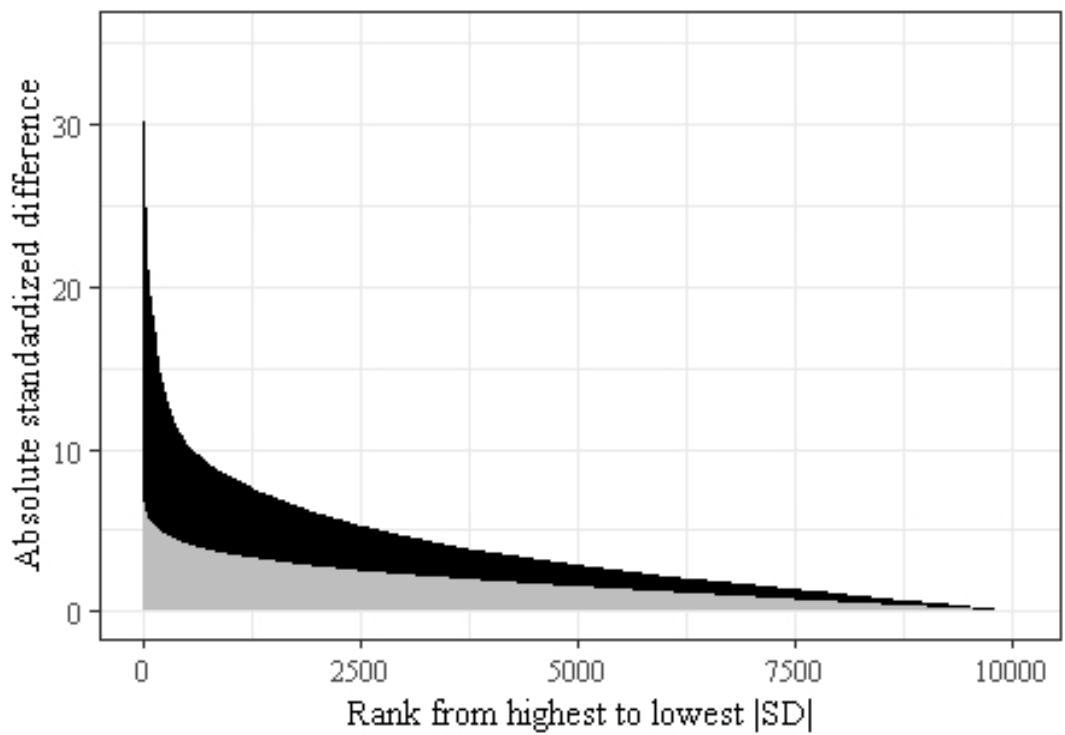

*Figure C.2.2: Balancing before and after adjustment for selection – multiple treatment*

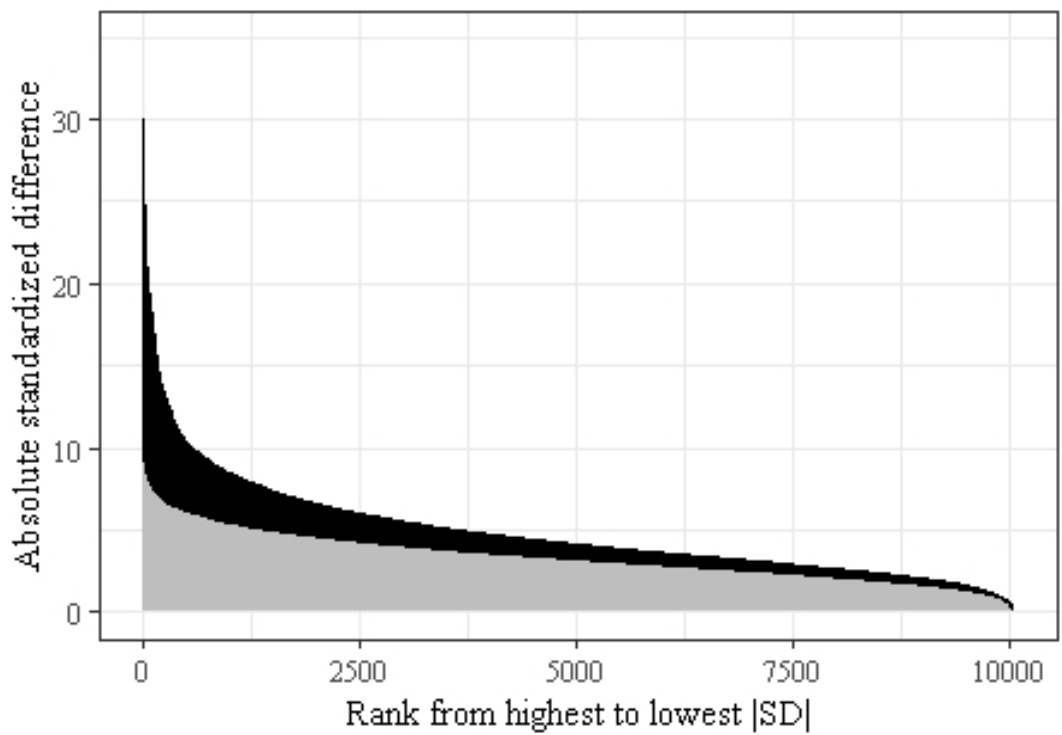

# Appendix D: Sensitivity analyses

## D.1 Sensitivity of results to penalty choice

Table D.1.1 shows in detail how important parameters of the analysis like the number of selected variables, implied weights and covariate balance vary for different penalty term choices. To get an overview, we report the mean values of each parameter over the 13 different outcome specifications.

The *number of variables* additionally selected for the predictions in the baseline is rather low. With nine and on average 4.5 for the propensity score in the binary and multiple treatment setting, respectively. The outcome predictions select on average 7.7 variables in the binary case and 4.6 in the multiple case. Combined with the 19 dummies, all specifications of the baseline use less than 30 variables for prediction. The number of selected variables varies substantially with the different penalty choice rules. The 1SE rule leads to very sparse models with zero to three additionally selected variables, while the 1SE+ and 2SE+ rules increase the selected variables up to 30.

The inspection of the implicit *weights* (see section 4.4) reveal no severe issues. The largest weight that one observation receives in percent of the total weights is on average at most 0.5% and always below 1%. This is far below the 4% threshold used to further trim observations, e.g., in Lechner and Strittmatter (2017). Another hint that the estimator produces no extreme weights is that the 10% largest weights make up for about 20% of the total weights indicating that the results are not driven by just a few observations. Another interesting exercise is to look at the number of negative weights that would indicate extrapolation. This is possible due to the global nature of Post-Lasso but seems not to be a big issue. Some observations receive negative weights, especially in the multiple treatment case. However, the largest negative weight is -3 and therefore less than 0.1% of the total weight.

*Table D.1.1: Selected variables, weights and balancing for different penalty term choices*

|  | Uncond. Diff. | Dummies | 1SE | **Min** | 1SE+ | 2SE+ |
|---|---|---|---|---|---|---|
|  | (1) | (2) | (3) | **(4)** | (5) | (6) |
| Binary treatment | | | | | | |
| *Number of selected variables:* | | | | | | |
| # of additional selected variables for treatment | - | - | 3 | **9** | 17 | 30 |
| Mean # of additional selected variables for outcomes | - | - | 1.4 | **7.7** | 17.5 | 24.6 |
| *Description of weights:* | | | | | | |
| Largest weight in % of total weights | - | 0.2 | 0.3 | **0.3** | 0.3 | 0.4 |
| Fraction of largest 10% weights of total weights in % | - | 12.5 | 17.7 | **18.2** | 19.2 | 19.8 |
| Number of negative weights | - | 0.0 | 0.5 | **2.8** | 5.6 | 6.1 |
| *Balancing of all 10,066 potential covariates:* | | | | | | |
| Maximum \|SD\| | 32.7 | 30.3 | 8.6 | **8.0** | 8.0 | 7.5 |
| Mean \|SD\| | 3.5 | 2.6 | 1.8 | **1.7** | 1.5 | 1.5 |
| Median \|SD\| | 2.6 | 1.9 | 1.6 | **1.5** | 1.3 | 1.3 |
| Fraction of variables with \|SD\| > 10 in % | 5.2 | 2.2 | 0.0 | **0.0** | 0.0 | 0.0 |
| Fraction of variables with \|SD\| > 5 in % | 25.1 | 12.3 | 4.4 | **2.5** | 1.2 | 1.0 |
| Multiple treatment | | | | | | |
| *Number of selected variables:* | | | | | | |
| Mean # of additionally selected variables for treatment | - | - | 1.0 | **4.5** | 14.8 | 27.5 |
| Mean # of additionally selected variables for outcomes | - | - | 0.9 | **4.6** | 11.2 | 17.7 |
| *Description of weights:* | | | | | | |
| Largest weight in % of total weights | - | 0.3 | 0.3 | **0.5** | 0.5 | 0.5 |
| Fraction of largest 10% weights of total weights in % | - | 13.9 | 15.9 | **18.2** | 19.8 | 21.0 |
| Number of negative weights | - | 6.0 | 17.2 | **24.6** | 76.5 | 108.9 |
| *Balancing of all 10,066 potential covariates:* | | | | | | |
| Maximum \|SD\| | 35.2 | 31.8 | 16.8 | **11.5** | 10.8 | 13.1 |
| Mean \|SD\| | 3.2 | 2.4 | 2.2 | **2.1** | 1.9 | 1.9 |
| Median \|SD\| | 2.4 | 1.9 | 1.9 | **1.7** | 1.6 | 1.6 |
| Fraction of variables with \|SD\| > 10 in % | 6.5 | 2.8 | 0.9 | **0.1** | 0.04 | 0.03 |
| Fraction of variables with \|SD\| > 5 in % | 38.3 | 24.5 | 20.1 | **16.2** | 12.2 | 12.6 |

Note: Table shows different characteristics of the estimations for different specifications. The numbers are averages over all outcomes. Column (1) shows the unconditional differences, column (2) the specification with state and school track dummies, and columns (3) to (6) the specifications obtained from the different penalty term choices. Column (4) marks the baseline. |SD| means absolute standardized difference.

Section 4.4 explains how we can use the calculated weights to assess *covariate balancing*. We start in the raw sample with standardized differences (SD) of potential confounders showing a maximum of over 30 as well as 25% / 38% (binary / multiple) having a SD larger than 5.[27] Just including state and school track dummies reduces the imbalances already substantially. Adding additional selected variables decreases the SD of all potential confounders further. The improvements in balancing are pronounced not only for the selected variables but for all potential controls. This indicates that the very sparse model is already sufficient to achieve reasonable balancing over all potential confounders. Penalty choices leading to larger models than the baseline reduce the fraction of potential confounders with absolute SD larger the 5 from 2.5% / 16% (binary / multiple) in the baseline to 1% / 12% for larger models.[28]

---
[27] Rosenbaum and Rubin (1985) consider values of above 20 as being large.

Table D.1.1 above shows that increasing model complexity leads to better balancing of potential confounders but comes at the cost of more extreme weights. Tables D.1.2 to D.1.7 below investigate whether larger models actually change the estimated effects. We observe a clear pattern that controlling for the state and school track dummies already decreases the observed unconditional differences to a large extent, which emphasizes that institutional knowledge is still important. Adding the additional selected controls drives the differences even more to zero, suggesting that the variable selection picks up important confounders. However, increasing the model complexity beyond the cross-validated minimum does not change the estimated coefficients dramatically. They vary from minimum to the 2SE+ rule specifications by at most one standard error of the baseline effects and the main qualitative conclusions remain valid. As expected from the discussion about more extreme weights, more complex models result also in slightly increased standard errors.

---

[28] Note that with more covariates than observations even methods that achieve perfect balancing of covariates (e.g., Graham, Pinto, & Egel, 2012, 2016; Hainmueller, 2012) could not find weights that perfectly balance all potential confounders.

*Table D.1.2: Average treatment effects of being musically active for different penalty choices*

|  | Uncond. Diff. | Dummies | 1SE | Minimum | 1SE+ | 2SE+ |
|---|---|---|---|---|---|---|
|  | (1) | (2) | (3) | (4) | (5) | (6) |
| *Cognitive skills (standardized)* | | | | | | |
| Science | 0.30*** | 0.11*** | 0.12*** | 0.11*** | 0.09*** | 0.09*** |
|  | (0.026) | (0.022) | (0.022) | (0.022) | (0.023) | (0.023) |
| Math | 0.23*** | 0.02 | 0.09*** | 0.08*** | 0.08*** | 0.07*** |
|  | (0.026) | (0.022) | (0.022) | (0.022) | (0.022) | (0.023) |
| Vocabulary | 0.29*** | 0.10*** | 0.12*** | 0.11*** | 0.09*** | 0.10*** |
|  | (0.026) | (0.024) | (0.024) | (0.024) | (0.025) | (0.024) |
| Reading | 0.20*** | 0.05** | -0.02 | -0.03 | -0.02 | -0.02 |
|  | (0.026) | (0.025) | (0.026) | (0.025) | (0.026) | (0.026) |
| ICT | 0.35*** | 0.15*** | 0.13*** | 0.12*** | 0.11*** | 0.11*** |
|  | (0.026) | (0.023) | (0.024) | (0.023) | (0.024) | (0.024) |
| *School performance (standardized)* | | | | | | |
| German grade | 0.34*** | 0.26*** | 0.14*** | 0.12*** | 0.10*** | 0.10*** |
|  | (0.026) | (0.026) | (0.028) | (0.028) | (0.030) | (0.029) |
| Math grade | 0.11*** | 0.07*** | 0.06** | 0.05* | 0.04 | 0.04 |
|  | (0.026) | (0.027) | (0.028) | (0.028) | (0.028) | (0.028) |
| Average grade German & math | 0.26*** | 0.19*** | 0.11*** | 0.09*** | 0.08*** | 0.08*** |
|  | (0.026) | (0.027) | (0.029) | (0.028) | (0.029) | (0.029) |
| *Big Five (standardized)* | | | | | | |
| Extraversion | 0.08*** | 0.07*** | 0.03 | 0.03 | 0.02 | 0.02 |
|  | (0.026) | (0.026) | (0.027) | (0.027) | (0.028) | (0.028) |
| Agreeableness | 0.13*** | 0.15*** | 0.11*** | 0.11*** | 0.11*** | 0.11*** |
|  | (0.026) | (0.027) | (0.028) | (0.028) | (0.029) | (0.029) |
| Conscientiousness | 0.03 | 0.07*** | -0.04 | -0.04 | -0.04 | -0.03 |
|  | (0.026) | (0.025) | (0.026) | (0.025) | (0.027) | (0.027) |
| Neuroticism | 0.07*** | 0.08*** | 0.000 | 0.001 | 0.005 | -0.002 |
|  | (0.026) | (0.026) | (0.028) | (0.028) | (0.030) | (0.030) |
| Openness | 0.47*** | 0.46*** | 0.33*** | 0.31*** | 0.29*** | 0.29*** |
|  | (0.025) | (0.026) | (0.027) | (0.028) | (0.028) | (0.028) |

Note: Column (1) shows unconditional mean differences between students who play at least one day of music compared to non-musicians, (2) shows the ATE only controlling for state and school track dummies, (3) – (6) show ATEs obtained for different penalty term choices for the Farrell (2015) estimator using Post-Lasso. All outcome variables are standardized to mean zero and variance one. Higher grades are better. State and school track dummies enter the selection unpenalized. Standard errors in brackets are clustered at school level. *, **, *** indicate statistical significance at the 10%, 5%, 1% level, respectively.

*Table D.1.3: Average treatment effects for Low vs. No with different penalty choices*

|  | Uncond. Diff. | Dummies | 1SE | Min | 1SE+ | 2SE+ |
|---|---|---|---|---|---|---|
| *Cognitive skills (standardized)* | | | | | | |
| Science | 0.12*** | 0.00 | 0.05 | 0.04 | 0.03 | 0.03 |
|  | (0.036) | (0.034) | (0.033) | (0.033) | (0.033) | (0.034) |
| Math | 0.09** | -0.04 | 0.06* | 0.05 | 0.05 | 0.04 |
|  | (0.037) | (0.032) | (0.031) | (0.031) | (0.031) | (0.032) |
| Vocabulary | 0.10*** | -0.02 | 0.03 | 0.02 | 0.01 | 0.02 |
|  | (0.037) | (0.033) | (0.032) | (0.031) | (0.030) | (0.029) |
| Reading | 0.17*** | 0.07** | 0.05 | 0.01 | 0.02 | 0.02 |
|  | (0.037) | (0.035) | (0.036) | (0.036) | (0.035) | (0.036) |
| ICT | 0.20*** | 0.07** | 0.05* | 0.06* | 0.08** | 0.07** |
|  | (0.036) | (0.032) | (0.032) | (0.033) | (0.032) | (0.031) |
| *School performance (standardized)* | | | | | | |
| German grade | 0.26*** | 0.22*** | 0.11*** | 0.11*** | 0.09** | 0.09** |
|  | (0.035) | (0.037) | (0.036) | (0.036) | (0.035) | (0.036) |
| Math grade | 0.05 | 0.03 | 0.03 | 0.04 | 0.03 | 0.01 |
|  | (0.036) | (0.038) | (0.039) | (0.039) | (0.039) | (0.041) |
| Average grade German & math | 0.17*** | 0.14*** | 0.10*** | 0.09** | 0.07* | 0.06 |
|  | (0.035) | (0.037) | (0.038) | (0.037) | (0.038) | (0.038) |
| *Big Five (standardized)* | | | | | | |
| Extraversion | 0.03 | 0.03 | 0.00 | 0.00 | -0.01 | -0.01 |
|  | (0.036) | (0.038) | (0.039) | (0.039) | (0.040) | (0.040) |
| Agreeableness | 0.14*** | 0.15*** | 0.14*** | 0.11*** | 0.10** | 0.10** |
|  | (0.036) | (0.037) | (0.038) | (0.038) | (0.038) | (0.039) |
| Conscientiousness | 0.02 | 0.06 | 0.00 | -0.04 | -0.04 | -0.03 |
|  | (0.036) | (0.037) | (0.037) | (0.037) | (0.037) | (0.039) |
| Neuroticism | 0.02 | 0.14*** | 0.04 | 0.05 | 0.07* | 0.06 |
|  | (0.036) | (0.038) | (0.038) | (0.039) | (0.038) | (0.040) |
| Openness | 0.14*** | 0.29*** | 0.16*** | 0.13*** | 0.13*** | 0.12*** |
|  | (0.036) | (0.038) | (0.037) | (0.038) | (0.037) | (0.038) |

Note: Column (1) shows unconditional mean, (2) shows the effects only controlling for state and school track dummies, (3) – (6) show obtained for different penalty term choices for the Farrell (2015) estimator using Post-Lasso. All outcome variables are standardized to mean zero and variance one. Higher grades are better. State and school track dummies enter the selection unpenalized. Standard errors in brackets are clustered at school level. *, **, *** indicate statistical significance at the 10%, 5%, 1% level, respectively.

*Table D.1.4: Average treatment effects for Medium vs. No with different penalty choices*

|  | Uncond. Diff. | Dummies | 1SE | Min | 1SE+ | 2SE+ |
|---|---|---|---|---|---|---|
| *Cognitive skills (standardized)* | | | | | | |
| Science | 0.34*** | 0.13*** | 0.14*** | 0.14*** | 0.13*** | 0.13*** |
|  | (0.033) | (0.030) | (0.030) | (0.030) | (0.030) | (0.030) |
| Math | 0.26*** | 0.04 | 0.09*** | 0.12*** | 0.09*** | 0.08*** |
|  | (0.033) | (0.029) | (0.029) | (0.029) | (0.029) | (0.029) |
| Vocabulary | 0.32*** | 0.13*** | 0.17*** | 0.16*** | 0.14*** | 0.15*** |
|  | (0.032) | (0.029) | (0.029) | (0.029) | (0.028) | (0.028) |
| Reading | 0.19*** | 0.03 | 0.00 | -0.04 | -0.04 | -0.07* |
|  | (0.033) | (0.032) | (0.032) | (0.032) | (0.034) | (0.035) |
| ICT | 0.38*** | 0.16*** | 0.15*** | 0.15*** | 0.15*** | 0.15*** |
|  | (0.033) | (0.029) | (0.029) | (0.030) | (0.031) | (0.031) |
| *School performance (standardized)* | | | | | | |
| German grade | 0.37*** | 0.28*** | 0.14*** | 0.13*** | 0.09** | 0.09** |
|  | (0.033) | (0.034) | (0.033) | (0.034) | (0.035) | (0.036) |
| Math grade | 0.14*** | 0.10*** | 0.09*** | 0.08** | 0.07** | 0.06 |
|  | (0.033) | (0.035) | (0.035) | (0.036) | (0.037) | (0.038) |
| Average grade German & math | 0.29*** | 0.22*** | 0.14*** | 0.13*** | 0.08** | 0.09** |
|  | (0.033) | (0.034) | (0.034) | (0.035) | (0.036) | (0.036) |
| *Big Five (standardized)* | | | | | | |
| Extraversion | 0.10*** | 0.08** | 0.06 | 0.04 | 0.06 | 0.04 |
|  | (0.033) | (0.035) | (0.036) | (0.036) | (0.038) | (0.040) |
| Agreeableness | 0.14*** | 0.15*** | 0.13*** | 0.10*** | 0.10*** | 0.09** |
|  | (0.033) | (0.035) | (0.035) | (0.036) | (0.038) | (0.039) |
| Conscientiousness | 0.02 | 0.06* | -0.06* | -0.06* | -0.06* | -0.07** |
|  | (0.033) | (0.034) | (0.034) | (0.034) | (0.035) | (0.036) |
| Neuroticism | 0.08** | 0.08** | -0.02 | -0.02 | 0.00 | 0.01 |
|  | (0.033) | (0.035) | (0.035) | (0.035) | (0.036) | (0.038) |
| Openness | 0.49*** | 0.47*** | 0.35*** | 0.33*** | 0.32*** | 0.33*** |
|  | (0.032) | (0.033) | (0.033) | (0.034) | (0.035) | (0.035) |

Note: Column (1) shows unconditional mean, (2) shows the effects only controlling for state and school track dummies, (3) – (6) show obtained for different penalty term choices for the Farrell (2015) estimator using Post-Lasso. All outcome variables are standardized to mean zero and variance one. Higher grades are better. State and school track dummies enter the selection unpenalized. Standard errors in brackets are clustered at school level. *, **, *** indicate statistical significance at the 10%, 5%, 1% level, respectively.

*Table D.1.4: Average treatment effects for High vs. No with different penalty choices*

|  | Uncond. Diff. | Dummies | 1SE | Min | 1SE+ | 2SE+ |
|---|---|---|---|---|---|---|
| *Cognitive skills (standardized)* | | | | | | |
| Science | 0.44*** | 0.22*** | 0.23*** | 0.17*** | 0.17*** | 0.17*** |
|  | (0.036) | (0.035) | (0.035) | (0.035) | (0.035) | (0.036) |
| Math | 0.34*** | 0.08** | 0.13*** | 0.10*** | 0.10*** | 0.10*** |
|  | (0.038) | (0.036) | (0.036) | (0.035) | (0.035) | (0.033) |
| Vocabulary | 0.34*** | 0.22*** | 0.23*** | 0.18*** | 0.17*** | 0.20*** |
|  | (0.036) | (0.033) | (0.033) | (0.034) | (0.033) | (0.032) |
| Reading | 0.24*** | 0.05 | 0.01 | -0.01 | -0.01 | -0.02 |
|  | (0.038) | (0.039) | (0.038) | (0.039) | (0.039) | (0.039) |
| ICT | 0.47*** | 0.23*** | 0.21*** | 0.18*** | 0.16*** | 0.16*** |
|  | (0.036) | (0.033) | (0.034) | (0.035) | (0.033) | (0.033) |
| *School performance (standardized)* | | | | | | |
| German grade | 0.39*** | 0.28*** | 0.21*** | 0.16*** | 0.13*** | 0.11*** |
|  | (0.039) | (0.042) | (0.042) | (0.042) | (0.042) | (0.040) |
| Math grade | 0.13*** | 0.07* | 0.07 | 0.04 | 0.01 | 0.02 |
|  | (0.040) | (0.043) | (0.044) | (0.045) | (0.045) | (0.045) |
| Average grade German & math | 0.30*** | 0.20*** | 0.16*** | 0.10** | 0.07 | 0.07 |
|  | (0.040) | (0.043) | (0.044) | (0.043) | (0.043) | (0.044) |
| *Big Five (standardized)* | | | | | | |
| Extraversion | 0.11*** | 0.10** | 0.08* | 0.07 | 0.06 | 0.05 |
|  | (0.039) | (0.043) | (0.044) | (0.045) | (0.045) | (0.045) |
| Agreeableness | 0.12*** | 0.14*** | 0.12*** | 0.11*** | 0.13*** | 0.12*** |
|  | (0.039) | (0.042) | (0.043) | (0.042) | (0.043) | (0.045) |
| Conscientiousness | 0.06 | 0.08* | 0.00 | -0.01 | -0.03 | -0.03 |
|  | (0.040) | (0.044) | (0.042) | (0.043) | (0.043) | (0.044) |
| Neuroticism | -0.02 | -0.02 | -0.07 | -0.04 | -0.02 | -0.03 |
|  | (0.039) | (0.042) | (0.042) | (0.044) | (0.044) | (0.046) |
| Openness | 0.62*** | 0.63*** | 0.54*** | 0.50*** | 0.50*** | 0.50*** |
|  | (0.037) | (0.041) | (0.041) | (0.043) | (0.043) | (0.043) |

Note: Column (1) shows unconditional mean, (2) shows the effects only controlling for state and school track dummies, (3) – (6) show obtained for different penalty term choices for the Farrell (2015) estimator using Post-Lasso. All outcome variables are standardized to mean zero and variance one. Higher grades are better. State and school track dummies enter the selection unpenalized. Standard errors in brackets are clustered at school level. *, **, *** indicate statistical significance at the 10%, 5%, 1% level, respectively.

*Table D.1.5: Average treatment effects for Medium vs. Low with different penalty choices*

|  | Uncond. Diff. | Dummies | 1SE | Min | 1SE+ | 2SE+ |
|---|---|---|---|---|---|---|
| *Cognitive skills (standardized)* | | | | | | |
| Science | 0.21*** | 0.13*** | 0.09** | 0.10*** | 0.10*** | 0.11*** |
|  | (0.041) | (0.038) | (0.037) | (0.037) | (0.037) | (0.037) |
| Math | 0.18*** | 0.08** | 0.03 | 0.07** | 0.05 | 0.04 |
|  | (0.042) | (0.036) | (0.035) | (0.036) | (0.035) | (0.034) |
| Vocabulary | 0.22*** | 0.14*** | 0.14*** | 0.13*** | 0.13*** | 0.13*** |
|  | (0.040) | (0.036) | (0.035) | (0.034) | (0.032) | (0.032) |
| Reading | 0.02 | -0.04 | -0.04 | -0.06 | -0.06 | -0.08** |
|  | (0.040) | (0.039) | (0.039) | (0.040) | (0.041) | (0.041) |
| ICT | 0.17*** | 0.09** | 0.09** | 0.09** | 0.08** | 0.08** |
|  | (0.040) | (0.036) | (0.036) | (0.037) | (0.037) | (0.037) |
| *School performance (standardized)* | | | | | | |
| German grade | 0.11*** | 0.06 | 0.03 | 0.03 | 0.00 | 0.00 |
|  | (0.040) | (0.042) | (0.040) | (0.041) | (0.040) | (0.042) |
| Math grade | 0.09** | 0.07 | 0.06 | 0.04 | 0.04 | 0.05 |
|  | (0.041) | (0.043) | (0.043) | (0.044) | (0.045) | (0.047) |
| Average grade German & math | 0.12*** | 0.08* | 0.03 | 0.03 | 0.01 | 0.03 |
|  | (0.041) | (0.043) | (0.042) | (0.043) | (0.043) | (0.043) |
| *Big Five (standardized)* | | | | | | |
| Extraversion | 0.06 | 0.06 | 0.06 | 0.04 | 0.07 | 0.05 |
|  | (0.041) | (0.043) | (0.043) | (0.044) | (0.046) | (0.047) |
| Agreeableness | -0.01 | 0.00 | 0.00 | -0.01 | 0.00 | -0.01 |
|  | (0.040) | (0.042) | (0.042) | (0.043) | (0.044) | (0.044) |
| Conscientiousness | -0.01 | 0.00 | -0.06 | -0.02 | -0.03 | -0.04 |
|  | (0.040) | (0.042) | (0.041) | (0.042) | (0.042) | (0.043) |
| Neuroticism | -0.07 | -0.06 | -0.07 | -0.07* | -0.08* | -0.06 |
|  | (0.041) | (0.043) | (0.042) | (0.043) | (0.043) | (0.044) |
| Openness | 0.19*** | 0.18*** | 0.19*** | 0.19*** | 0.19*** | 0.21*** |
|  | (0.040) | (0.042) | (0.041) | (0.042) | (0.041) | (0.042) |

Note: Column (1) shows unconditional mean, (2) shows the effects only controlling for state and school track dummies, (3) – (6) show obtained for different penalty term choices for the Farrell (2015) estimator using Post-Lasso. All outcome variables are standardized to mean zero and variance one. Higher grades are better. State and school track dummies enter the selection unpenalized. Standard errors in brackets are clustered at school level. *, **, *** indicate statistical significance at the 10%, 5%, 1% level, respectively.

*Table D.1.6: Average treatment effects for High vs. Low with different penalty choices*

|  | Uncond. Diff. | Dummies | 1SE | Min | 1SE+ | 2SE+ |
|---|---|---|---|---|---|---|
| *Cognitive skills (standardized)* | | | | | | |
| Science | 0.31*** | 0.22*** | 0.19*** | 0.13*** | 0.15*** | 0.15*** |
|  | (0.043) | (0.042) | (0.041) | (0.042) | (0.041) | (0.042) |
| Math | 0.26*** | 0.13*** | 0.07* | 0.05 | 0.05 | 0.07* |
|  | (0.046) | (0.042) | (0.041) | (0.040) | (0.040) | (0.038) |
| Vocabulary | 0.34*** | 0.23*** | 0.20*** | 0.15*** | 0.16*** | 0.18*** |
|  | (0.043) | (0.040) | (0.038) | (0.038) | (0.036) | (0.035) |
| Reading | 0.07 | -0.03 | -0.04 | -0.02 | -0.02 | -0.04 |
|  | (0.045) | (0.045) | (0.044) | (0.045) | (0.045) | (0.045) |
| ICT | 0.27*** | 0.16*** | 0.16*** | 0.11*** | 0.09** | 0.09** |
|  | (0.042) | (0.039) | (0.039) | (0.041) | (0.038) | (0.038) |
| *School performance (standardized)* | | | | | | |
| German grade | 0.13*** | 0.07 | 0.10** | 0.05 | 0.05 | 0.02 |
|  | (0.045) | (0.049) | (0.048) | (0.048) | (0.047) | (0.045) |
| Math grade | 0.08* | 0.04 | 0.04 | 0.00 | -0.02 | 0.01 |
|  | (0.047) | (0.050) | (0.050) | (0.052) | (0.051) | (0.053) |
| Average grade German & math | 0.12*** | 0.06 | 0.06 | 0.01 | 0.00 | 0.01 |
|  | (0.047) | (0.050) | (0.050) | (0.050) | (0.050) | (0.050) |
| *Big Five (standardized)* | | | | | | |
| Extraversion | 0.08* | 0.08 | 0.08 | 0.07 | 0.07 | 0.06 |
|  | (0.046) | (0.050) | (0.050) | (0.052) | (0.052) | (0.051) |
| Agreeableness | -0.02 | -0.01 | -0.01 | 0.01 | 0.03 | 0.02 |
|  | (0.045) | (0.049) | (0.049) | (0.049) | (0.049) | (0.050) |
| Conscientiousness | 0.03 | 0.02 | 0.01 | 0.04 | 0.01 | 0.00 |
|  | (0.046) | (0.050) | (0.048) | (0.049) | (0.048) | (0.050) |
| Neuroticism | -0.16*** | -0.16*** | -0.11** | -0.10* | -0.09* | -0.09* |
|  | (0.046) | (0.049) | (0.049) | (0.050) | (0.050) | (0.051) |
| Openness | 0.33*** | 0.34*** | 0.38*** | 0.37*** | 0.37*** | 0.38*** |
|  | (0.044) | (0.048) | (0.048) | (0.049) | (0.048) | (0.048) |

Note: Column (1) shows unconditional mean, (2) shows the effects only controlling for state and school track dummies, (3) – (6) show obtained for different penalty term choices for the Farrell (2015) estimator using Post-Lasso. All outcome variables are standardized to mean zero and variance one. Higher grades are better. State and school track dummies enter the selection unpenalized. Standard errors in brackets are clustered at school level. *, **, *** indicate statistical significance at the 10%, 5%, 1% level, respectively.

*Table D.1.7: Average treatment effects for High vs. Medium with different penalty choices*

|  | Uncond. Diff. | Dummies | 1SE | Min | 1SE+ | 2SE+ |
|---|---|---|---|---|---|---|
| *Cognitive skills (standardized)* | | | | | | |
| Science | 0.10** | 0.09** | 0.10** | 0.03 | 0.04 | 0.04 |
|  | (0.041) | (0.039) | (0.039) | (0.039) | (0.038) | (0.038) |
| Math | 0.08* | 0.04 | 0.04 | -0.02 | 0.00 | 0.02 |
|  | (0.043) | (0.040) | (0.040) | (0.038) | (0.038) | (0.035) |
| Vocabulary | 0.12*** | 0.09** | 0.06* | 0.02 | 0.03 | 0.05 |
|  | (0.039) | (0.037) | (0.035) | (0.036) | (0.035) | (0.034) |
| Reading | 0.05 | 0.02 | 0.00 | 0.03 | 0.04 | 0.05 |
|  | (0.042) | (0.043) | (0.042) | (0.042) | (0.044) | (0.044) |
| ICT | 0.09** | 0.07* | 0.07* | 0.03 | 0.01 | 0.01 |
|  | (0.040) | (0.037) | (0.037) | (0.039) | (0.037) | (0.038) |
| *School performance (standardized)* | | | | | | |
| German grade | 0.02 | 0.00 | 0.07 | 0.03 | 0.04 | 0.02 |
|  | (0.044) | (0.046) | (0.046) | (0.046) | (0.046) | (0.045) |
| Math grade | -0.01 | -0.03 | -0.02 | -0.04 | -0.06 | -0.04 |
|  | (0.044) | (0.048) | (0.048) | (0.050) | (0.049) | (0.050) |
| Average grade German & math | 0.01 | -0.02 | 0.03 | -0.03 | -0.02 | -0.02 |
|  | (0.045) | (0.048) | (0.047) | (0.048) | (0.048) | (0.048) |
| *Big Five (standardized)* | | | | | | |
| Extraversion | 0.01 | 0.02 | 0.02 | 0.02 | 0.00 | 0.01 |
|  | (0.044) | (0.048) | (0.048) | (0.050) | (0.051) | (0.051) |
| Agreeableness | -0.01 | -0.01 | -0.01 | 0.01 | 0.02 | 0.03 |
|  | (0.043) | (0.047) | (0.047) | (0.047) | (0.048) | (0.049) |
| Conscientiousness | 0.03 | 0.02 | 0.07 | 0.05 | 0.03 | 0.04 |
|  | (0.044) | (0.048) | (0.046) | (0.047) | (0.047) | (0.047) |
| Neuroticism | -0.09** | -0.10** | -0.05 | -0.02 | -0.02 | -0.04 |
|  | (0.043) | (0.046) | (0.046) | (0.048) | (0.049) | (0.049) |
| Openness | 0.14*** | 0.16*** | 0.19*** | 0.18*** | 0.18*** | 0.17*** |
|  | (0.041) | (0.045) | (0.045) | (0.047) | (0.046) | (0.046) |

Note: Column (1) shows unconditional mean, (2) shows the effects only controlling for state and school track dummies, (3) – (6) show obtained for different penalty term choices for the Farrell (2015) estimator using Post-Lasso. All outcome variables are standardized to mean zero and variance one. Higher grades are better. State and school track dummies enter the selection unpenalized. Standard errors in brackets are clustered at school level. *, **, *** indicate statistical significance at the 10%, 5%, 1% level, respectively.

## D.2 Anatomy of double machine learning weights

Table D.2.1 reports the average correlations between the different components of equation (8). It shows that the DML weights ($w_t$) in this application are mostly driven by IPW weights ($w_t^p$). DML and IPW weights are highly correlated with an average correlation of 0.99 for the binary and the multiple treatment case. However, also the RA weights ($w_t^Y$) show high correlations with DML weights with 0.77 and 0.89 for the binary and the multiple treatment case, respectively. The explanation is that IPW and RA mostly agree on how to weight the outcomes to estimate the causal effect with correlations of 0.76 and 0.87 for the binary and the multiple treatment case, respectively. The adjustment weight ($w_t^{pY}$) that is subtracted is highly correlated with the RA weights (0.99).

*Table D.2.1: Average correlation of DML weights and its components*

|  | Binary | | | | Multiple | | | |
| --- | --- | --- | --- | --- | --- | --- | --- | --- |
|  | $w_t$ | $w_t^p$ | $w_t^Y$ | $w_t^{pY}$ | $w_t$ | $w_t^p$ | $w_t^Y$ | $w_t^{pY}$ |
| $w_t$ | 1.00 | | | | 1.00 | | | |
| $w_t^p$ | 0.99 | 1.00 | | | 0.99 | 1.00 | | |
| $w_t^Y$ | 0.77 | 0.76 | 1.00 | | 0.89 | 0.87 | 1.00 | |
| $w_t^{pY}$ | 0.76 | 0.77 | 0.99 | 1.00 | 0.88 | 0.88 | 0.99 | 1.00 |

Note: This table shows the average correlations between the different weights of equation (8) over all 13 outcomes shown in Table 2. The results are obtained by applying the Farrell (2015) estimator using Post-Lasso with penalty chosen at the minimum of 10-fold cross-validated MSE.

In a second step, we investigate the balancing properties of DML, IPW and RA separately in Table D.2.2. As expected from the correlations above, the balancing of DML and IPW weights are nearly identical. However, the DML balancing is for most indicators slightly better. In contrast, balancing of RA is substantially worse. This shows that DML has good reasons to rely mostly on IPW weights and to offset the influence of RA weights by subtracting it with the adjustment weights.

*Table D.2.2: Balancing of DML, IPW and RA weights*

|  | Binary | | | Multiple | | |
| --- | --- | --- | --- | --- | --- | --- |
|  | $w_t$ | $w_t^p$ | $w_t^Y$ | $w_t$ | $w_t^p$ | $w_t^Y$ |
| Maximum \|SD\| | 8.0 | 8.2 | 19.9 | 11.5 | 11.1 | 23.2 |
| Mean \|SD\| | 1.7 | 1.8 | 2.4 | 2.0 | 2.1 | 2.3 |
| Median \|SD\| | 1.5 | 1.6 | 1.9 | 2.1 | 2.1 | 2.2 |
| Fraction of variables with \|SD\| > 10 in % | 0.0 | 0.0 | 1.0 | 0.1 | 0.1 | 1.5 |
| Fraction of variables with \|SD\| > 5 in % | 2.5 | 2.8 | 9.2 | 16.2 | 17.0 | 21.8 |

Note: This table compares the average balancing results obtained from the different weights of equation (8) over all 13 outcomes shown in Table 2. The results are obtained by applying the Farrell (2015) estimator using Post-Lasso with penalty chosen at the minimum of 10-fold cross-validated MSE. |SD| means absolute standardized difference.

# D.3: Further sensitivity analyses

### D.3.1 Potentially endogenous controls

Section 4.1 discusses the concern that some of the available controls in the NEPS data might be themselves outcomes and are therefore excluded from the main specification. Table D.3.1 shows the results if those variables are included in the set of potential variables. Besides the marginally significant effect in math grades becoming insignificant, no striking differences relative to the main results in Table 2 are observed.

*Table D.3.1: Results including potentially endogenous control variables*

|  | Binary | Multiple | | | | | |
|---|---|---|---|---|---|---|---|
|  | Any - No | Low – No | Med - No | High - No | Med - Low | High - Low | High - Med |
| *Cognitive Skills (standardized)* | | | | | | | |
| Science | 0.10*** | 0.04 | 0.13*** | 0.18*** | 0.09** | 0.13*** | 0.04 |
|  | (0.02) | (0.03) | (0.03) | (0.04) | (0.04) | (0.04) | (0.04) |
| Math | 0.07*** | 0.05 | 0.10*** | 0.09*** | 0.05 | 0.04 | -0.01 |
|  | (0.02) | (0.03) | (0.03) | (0.03) | (0.04) | (0.04) | (0.04) |
| Vocabulary | 0.10*** | 0.01 | 0.16*** | 0.18*** | 0.14*** | 0.17*** | 0.02 |
|  | (0.02) | (0.03) | (0.03) | (0.03) | (0.03) | (0.04) | (0.04) |
| Reading | -0.02 | 0.02 | -0.04 | 0.00 | -0.06 | -0.02 | 0.04 |
|  | (0.02) | (0.04) | (0.03) | (0.04) | (0.04) | (0.05) | (0.04) |
| ICT | 0.12*** | 0.06** | 0.15*** | 0.17*** | 0.09*** | 0.11*** | 0.02 |
|  | (0.02) | (0.03) | (0.03) | (0.03) | (0.04) | (0.04) | (0.04) |
| *School performance (standardized)* | | | | | | | |
| German grade | 0.11*** | 0.09*** | 0.12*** | 0.15*** | 0.03 | 0.06 | 0.03 |
|  | (0.03) | (0.03) | (0.03) | (0.04) | (0.04) | (0.05) | (0.04) |
| Math grade | 0.02 | 0.01 | 0.05 | 0.02 | 0.03 | 0.01 | -0.03 |
|  | (0.03) | (0.04) | (0.03) | (0.04) | (0.04) | (0.05) | (0.05) |
| Average grade German & math | 0.06** | 0.06 | 0.07** | 0.08** | 0.02 | 0.03 | 0.01 |
|  | (0.03) | (0.04) | (0.03) | (0.04) | (0.04) | (0.05) | (0.04) |
| *Big Five (standardized)* | | | | | | | |
| Extraversion | 0.03 | 0.00 | 0.04 | 0.07 | 0.04 | 0.07 | 0.03 |
|  | (0.03) | (0.04) | (0.04) | (0.05) | (0.04) | (0.05) | (0.05) |
| Agreeableness | 0.09*** | 0.10*** | 0.08** | 0.12*** | -0.02 | 0.02 | 0.04 |
|  | (0.03) | (0.04) | (0.04) | (0.04) | (0.04) | (0.05) | (0.05) |
| Conscientiousness | -0.05** | -0.05 | -0.06* | -0.01 | -0.01 | 0.04 | 0.05 |
|  | (0.03) | (0.04) | (0.03) | (0.04) | (0.04) | (0.05) | (0.05) |
| Neuroticism | -0.01 | 0.05 | -0.03 | -0.03 | -0.07* | -0.08 | -0.01 |
|  | (0.03) | (0.04) | (0.04) | (0.05) | (0.04) | (0.05) | (0.05) |
| Openness | 0.31*** | 0.14*** | 0.32*** | 0.50*** | 0.18*** | 0.36*** | 0.18*** |
|  | (0.03) | (0.04) | (0.03) | (0.04) | (0.04) | (0.05) | (0.05) |

Note: This table shows the estimated effects comparing different intensities of musical practice. All outcome variables are standardized to mean zero and variance one. Higher grades are better. The results are obtained by applying the Farrell (2015) estimator using Post-Lasso with penalty chosen at the minimum of 10-fold cross-validated MSE. State and school track dummies enter the selection unpenalized. Standard errors in brackets are clustered at school level. *, **, *** indicate statistical significance at the 10%, 5%, 1% level, respectively.

D.3.2 Different control groups

We check the sensitivity of the results to the choice of the group of non-musicians. Table D.3.2.1 shows the results using all students, not only extracurricularly active students. Table D.3.2.2 compares musicians to non-musicians doing sports as in Cabane et al. (2016).

We find some minor differences for the full sample approach in Table 2.2.1. The effects for grades are larger and especially for math grades highly significant. Further, extraversion now shows significantly positive effects. There are two potential explanations for this observation. (i) We are not able to control for the first selection step into being active at all, because we do not observe early personality traits that lead to extracurricular activities. This seems to be a valid concern as early extraversion could be a main driver into extracurricular activities and is not sufficiently controlled when including completely inactive students. Therefore, following the arguments of Cabane et al. (2016) and excluding inactive students seems to be crucial. (ii) Heterogeneous effects could also lead to differences. This would mean that students without any extracurricular activities have substantially higher positive effects of music on grades and extraversion.

Restricting the control group of non-musicians to sporty students in Table D.3.2.2 changes the estimated effects only marginally.

*Table D.3.2.1: Results using all students in the analysis*

|  | Binary | Multiple | | | | | |
|---|---|---|---|---|---|---|---|
|  | Any - No | Low - No | Med - No | High - No | Med - Low | High - Low | High - Med |
| *Cognitive Skills (standardized)* | | | | | | | |
| Science | 0.10*** | 0.03 | 0.13*** | 0.17*** | 0.10*** | 0.14*** | 0.04 |
|  | (0.02) | (0.03) | (0.03) | (0.03) | (0.04) | (0.04) | (0.04) |
| Math | 0.09*** | 0.08*** | 0.14*** | 0.14*** | 0.06* | 0.06 | 0.00 |
|  | (0.02) | (0.03) | (0.03) | (0.03) | (0.03) | (0.04) | (0.04) |
| Vocabulary | 0.09*** | 0.01 | 0.15*** | 0.17*** | 0.14*** | 0.16*** | 0.02 |
|  | (0.02) | (0.03) | (0.03) | (0.03) | (0.03) | (0.04) | (0.04) |
| Reading | -0.02 | 0.02 | -0.04 | -0.02 | -0.06 | -0.03 | 0.02 |
|  | (0.02) | (0.03) | (0.03) | (0.04) | (0.04) | (0.04) | (0.04) |
| ICT | 0.09*** | 0.04 | 0.13*** | 0.17*** | 0.09** | 0.13*** | 0.04 |
|  | (0.02) | (0.03) | (0.03) | (0.03) | (0.04) | (0.04) | (0.04) |
| *School performance (standardized)* | | | | | | | |
| German grade | 0.15*** | 0.14*** | 0.16*** | 0.17*** | 0.02 | 0.03 | 0.01 |
|  | (0.03) | (0.03) | (0.03) | (0.04) | (0.04) | (0.05) | (0.04) |
| Math grade | 0.10*** | 0.10*** | 0.13*** | 0.08* | 0.03 | -0.02 | -0.05 |
|  | (0.03) | (0.04) | (0.03) | (0.04) | (0.04) | (0.05) | (0.05) |
| Average grade German & math | 0.14*** | 0.15*** | 0.18*** | 0.12*** | 0.02 | -0.04 | -0.06 |
|  | (0.03) | (0.03) | (0.03) | (0.04) | (0.04) | (0.05) | (0.05) |
| *Big Five (standardized)* | | | | | | | |
| Extraversion | 0.06*** | 0.05 | 0.08** | 0.09** | 0.03 | 0.04 | 0.01 |
|  | (0.02) | (0.04) | (0.03) | (0.04) | (0.04) | (0.05) | (0.05) |
| Agreeableness | 0.13*** | 0.12*** | 0.11*** | 0.14*** | -0.01 | 0.01 | 0.02 |
|  | (0.03) | (0.04) | (0.03) | (0.04) | (0.04) | (0.05) | (0.05) |
| Conscientiousness | 0.01 | 0.01 | -0.02 | 0.03 | -0.03 | 0.03 | 0.06 |
|  | (0.02) | (0.03) | (0.03) | (0.04) | (0.04) | (0.05) | (0.05) |
| Neuroticism | -0.01 | 0.04 | -0.04 | -0.05 | -0.07* | -0.08* | -0.01 |
|  | (0.03) | (0.03) | (0.03) | (0.04) | (0.04) | (0.05) | (0.05) |
| Openness | 0.30*** | 0.14*** | 0.33*** | 0.49*** | 0.19*** | 0.35*** | 0.17*** |
|  | (0.03) | (0.03) | (0.03) | (0.04) | (0.04) | (0.05) | (0.05) |
| (Mean) # of selected variables for treatment | 21 | 7.0 | | | | | |
| Mean # of selected variables for outcomes | 10.5 | 5.9 | | | | | |
| # of observations trimmed | 12 | 76 | | | | | |

Note: This table shows the estimated effects comparing different intensities of musical practice using 6,898 students. All outcome variables are standardized to mean zero and variance one. Higher grades are better. The results are obtained by applying the Farrell (2015) estimator using Post-Lasso with penalty chosen at the minimum of 10-fold cross-validated MSE. State and school track dummies enter the selection unpenalized. Standard errors in brackets are clustered at school level. *, **, *** indicate statistical significance at the 10%, 5%, 1% level, respectively.

*Table D.3.2.2: Results using only children being active in sports as control group*

|  | Binary | Multiple | | | | | |
|---|---|---|---|---|---|---|---|
|  | Any - No | Low - No | Med - No | High - No | Med - Low | High - Low | High - Med |
| *Cognitive Skills (standardized)* | | | | | | | |
| Science | 0.13*** | 0.06* | 0.18*** | 0.20*** | 0.12*** | 0.14*** | 0.02 |
|  | (0.02) | (0.03) | (0.03) | (0.04) | (0.04) | (0.04) | (0.04) |
| Math | 0.07** | 0.05 | 0.12*** | 0.10*** | 0.07* | 0.05 | -0.02 |
|  | (0.02) | (0.03) | (0.03) | (0.03) | (0.04) | (0.04) | (0.04) |
| Vocabulary | 0.13*** | 0.03 | 0.17*** | 0.19*** | 0.13*** | 0.16*** | 0.02 |
|  | (0.02) | (0.03) | (0.03) | (0.03) | (0.03) | (0.04) | (0.04) |
| Reading | -0.03 | 0.01 | -0.04 | -0.02*** | -0.05 | -0.03 | 0.02 |
|  | (0.03) | (0.04) | (0.03) | (0.04) | (0.04) | (0.04) | (0.04) |
| ICT | 0.14*** | 0.08*** | 0.18*** | 0.19*** | 0.09** | 0.11*** | 0.01 |
|  | (0.02) | (0.03) | (0.03) | (0.03) | (0.04) | (0.04) | (0.04) |
| *School performance (standardized)* | | | | | | | |
| German grade | 0.11*** | 0.09** | 0.12*** | 0.15*** | 0.03 | 0.06 | 0.03 |
|  | (0.03) | (0.04) | (0.03) | (0.04) | (0.04) | (0.05) | (0.05) |
| Math grade | 0.04 | 0.02 | 0.00 | 0.02 | 0.05 | 0.00 | -0.05 |
|  | (0.03) | (0.04) | (0.00) | (0.04) | (0.04) | (0.05) | (0.05) |
| Average grade German & math | 0.08** | 0.00 | 0.00 | 0.08* | 0.04 | 0.00 | -0.03 |
|  | (0.03) | (0.00) | (0.00) | (0.04) | (0.04) | (0.05) | (0.05) |
| *Big Five (standardized)* | | | | | | | |
| Extraversion | 0.00 | -0.02 | 0.02 | 0.05 | 0.04 | 0.07 | 0.03 |
|  | (0.03) | (0.04) | (0.04) | (0.05) | (0.04) | (0.05) | (0.05) |
| Agreeableness | 0.12*** | 0.12*** | 0.11*** | 0.12*** | -0.01 | 0.00 | 0.01 |
|  | (0.03) | (0.04) | (0.04) | (0.04) | (0.04) | (0.05) | (0.05) |
| Conscientiousness | -0.04 | -0.04 | -0.05 | 0.00 | -0.01 | 0.04 | 0.05 |
|  | (0.03) | (0.04) | (0.04) | (0.04) | (0.04) | (0.05) | (0.05) |
| Neuroticism | 0.01 | 0.05 | -0.02 | -0.04 | -0.07** | -0.10* | -0.02 |
|  | (0.03) | (0.04) | (0.04) | (0.04) | (0.04) | (0.05) | (0.05) |
| Openness | 0.33*** | 0.15*** | 0.35*** | 0.52*** | 0.19*** | 0.36*** | 0.17*** |
|  | (0.03) | (0.04) | (0.03) | (0.04) | (0.04) | (0.05) | (0.05) |

Note: This table shows the estimated effects comparing different intensities of musical practice using 5,611 students. All outcome variables are standardized to mean zero and variance one. Higher grades are better. The results are obtained by applying the Farrell (2015) estimator using Post-Lasso with penalty chosen at the minimum of 10-fold cross-validated MSE. State and school track dummies enter the selection unpenalized. Standard errors in brackets are clustered at school level. *, **, *** indicate statistical significance at the 10%, 5%, 1% level, respectively.

### D.3.3 Different common support enforcement

This section investigates the sensitivity of the results to different common support adjustments compared to the baseline rule explained in section E.2. Table D.3.3.1 shows only minor changes when no common support adjustment is carried out at all. Also trimming more aggressively than the baseline minimum / maximum rule leaves the results nearly unchanged. Table D.3.3.2 provides according results for trimming at the highest 1$^{st}$ and lowest 99$^{th}$ percentile of the propensity scores. This leads to overall trimming of more than 10% in the multiple treatment case.

*Table D.3.3.1: Main results without enforcing common support*

|  | Binary | Multiple | | | | | |
|---|---|---|---|---|---|---|---|
|  | Any - No | Low - No | Med - No | High - No | Med - Low | High - Low | High - Med |
| Cognitive Skills (standardized) | | | | | | | |
| Science | 0.10*** | 0.03 | 0.13*** | 0.17*** | 0.10*** | 0.14*** | 0.04 |
|  | (0.02) | (0.03) | (0.03) | (0.03) | (0.04) | (0.04) | (0.04) |
| Math | 0.08*** | 0.05* | 0.12*** | 0.10*** | 0.07* | 0.05 | -0.01 |
|  | (0.02) | (0.03) | (0.03) | (0.03) | (0.03) | (0.04) | (0.04) |
| Vocabulary | 0.11*** | 0.01 | 0.15*** | 0.19*** | 0.14*** | 0.17*** | 0.03 |
|  | (0.02) | (0.03) | (0.03) | (0.03) | (0.03) | (0.04) | (0.03) |
| Reading | -0.03 | 0.01 | -0.05 | -0.02 | -0.06 | -0.03 | 0.03 |
|  | (0.03) | (0.03) | (0.03) | (0.04) | (0.04) | (0.04) | (0.04) |
| ICT | 0.11*** | 0.06* | 0.15*** | 0.18*** | 0.09*** | 0.12*** | 0.02 |
|  | (0.02) | (0.03) | (0.03) | (0.03) | (0.04) | (0.04) | (0.04) |
| *School performance (standardized)* | | | | | | | |
| German grade | 0.12*** | 0.10*** | 0.13*** | 0.15*** | 0.03 | 0.05 | 0.02 |
|  | (0.03) | (0.03) | (0.03) | (0.04) | (0.04) | (0.05) | (0.04) |
| Math grade | 0.05* | 0.13*** | 0.07** | 0.03 | 0.04 | -0.01 | -0.05 |
|  | (0.03) | (0.03) | (0.04) | (0.04) | (0.04) | (0.05) | (0.05) |
| Average grade German & math | 0.09*** | 0.09** | 0.10*** | 0.08* | 0.02 | -0.01 | -0.02 |
|  | (0.03) | (0.04) | (0.03) | (0.04) | (0.04) | (0.05) | (0.04) |
| *Big Five (standardized)* | | | | | | | |
| Extraversion | 0.03 | 0.01 | 0.06 | 0.06 | 0.05 | 0.05 | 0.00 |
|  | (0.03) | (0.04) | (0.04) | (0.04) | (0.04) | (0.05) | (0.05) |
| Agreeableness | 0.11*** | 0.10*** | 0.10*** | 0.11*** | 0.00 | 0.01 | 0.01 |
|  | (0.03) | (0.04) | (0.04) | (0.04) | (0.04) | (0.05) | (0.05) |
| Conscientiousness | -0.04 | -0.05 | -0.06* | -0.01 | -0.01 | 0.04 | 0.05 |
|  | (0.03) | (0.04) | (0.03) | (0.04) | (0.04) | (0.05) | (0.05) |
| Neuroticism | 0.00 | 0.05 | -0.02 | -0.04 | -0.07* | -0.09* | -0.02 |
|  | (0.03) | (0.04) | (0.04) | (0.04) | (0.04) | (0.05) | (0.05) |
| Openness | 0.31*** | 0.14*** | 0.33*** | 0.50*** | 0.19*** | 0.36*** | 0.17*** |
|  | (0.03) | (0.04) | (0.03) | (0.04) | (0.04) | (0.05) | (0.04) |
| (Mean) # of selected variables for treatment | 9 | | | | 4.5 | | |
| Mean # of selected variables for outcomes | 9.4 | | | | 4.7 | | |
| # of observations trimmed | 0 | | | | 0 | | |

Note: This table shows the estimated effects comparing different intensities of musical practice. All outcome variables are standardized to mean zero and variance one. Higher grades are better. The results are obtained by applying the Farrell (2015) estimator using Post-Lasso with penalty chosen at the minimum of 10-fold cross-validated MSE. State and school track dummies enter the selection unpenalized. Standard errors in brackets are clustered at school level. *, **, *** indicate statistical significance at the 10%, 5%, 1% level, respectively.

*Table D.3.3.2: Main results with trimming at the 1st and 99th percentile*

|  | Binary | Multiple | | | | | |
|---|---|---|---|---|---|---|---|
|  | Any - No | Low - No | Med - No | High - No | Med - Low | High - Low | High - Med |
| *Cognitive Skills (standardized)* | | | | | | | |
| Science | 0.11*** | 0.04 | 0.15*** | 0.17*** | 0.11*** | 0.13*** | 0.02 |
|  | (0.02) | (0.03) | (0.03) | (0.03) | (0.04) | (0.04) | (0.04) |
| Math | 0.09*** | 0.05* | 0.14*** | 0.10*** | 0.09*** | 0.05 | -0.05 |
|  | (0.02) | (0.03) | (0.03) | (0.03) | (0.03) | (0.04) | (0.04) |
| Vocabulary | 0.11*** | 0.01 | 0.15*** | 0.18*** | 0.14*** | 0.17*** | 0.03 |
|  | (0.02) | (0.03) | (0.03) | (0.03) | (0.03) | (0.04) | (0.04) |
| Reading | -0.03 | 0.00 | -0.04 | 0.00 | -0.05 | 0.00 | 0.04 |
|  | (0.02) | (0.04) | (0.03) | (0.04) | (0.04) | (0.04) | (0.04) |
| ICT | 0.13*** | 0.05* | 0.17*** | 0.18*** | 0.11*** | 0.12*** | 0.01 |
|  | (0.02) | (0.03) | (0.03) | (0.03) | (0.04) | (0.04) | (0.04) |
| *School performance (standardized)* | | | | | | | |
| German grade | 0.11*** | 0.10*** | 0.12*** | 0.16*** | 0.02 | 0.06 | 0.04 |
|  | (0.03) | (0.04) | (0.03) | (0.04) | (0.04) | (0.05) | (0.04) |
| Math grade | 0.06** | 0.04 | 0.08** | 0.03 | 0.04 | 0.00 | -0.05 |
|  | (0.03) | (0.04) | (0.03) | (0.04) | (0.04) | (0.05) | (0.05) |
| Average grade German & math | 0.09*** | 0.08** | 0.12*** | 0.09** | 0.03 | 0.01 | -0.03 |
|  | (0.03) | (0.04) | (0.03) | (0.04) | (0.04) | (0.05) | (0.04) |
| *Big Five (standardized)* | | | | | | | |
| Extraversion | 0.02 | 0.02 | 0.07* | 0.07* | 0.05 | 0.05 | 0.00 |
|  | (0.03) | (0.04) | (0.04) | (0.04) | (0.04) | (0.05) | (0.05) |
| Agreeableness | 0.10*** | 0.10*** | 0.10*** | 0.11*** | 0.00 | 0.01 | 0.01 |
|  | (0.03) | (0.04) | (0.04) | (0.04) | (0.04) | (0.05) | (0.05) |
| Conscientiousness | -0.04* | -0.06* | -0.08** | 0.01 | -0.01 | 0.07 | 0.09* |
|  | (0.03) | (0.04) | (0.03) | (0.04) | (0.04) | (0.05) | (0.05) |
| Neuroticism | 0.01 | 0.06* | -0.03 | -0.04 | -0.09** | -0.11** | -0.02 |
|  | (0.03) | (0.04) | (0.03) | (0.04) | (0.04) | (0.05) | (0.05) |
| Openness | 0.32*** | 0.16*** | 0.37*** | 0.52*** | 0.21*** | 0.36*** | 0.15*** |
|  | (0.03) | (0.04) | (0.03) | (0.04) | (0.04) | (0.05) | (0.04) |
| (Mean) # of selected variables for treatment | 9 | 4.5 | | | | | |
| Mean # of selected variables for outcomes | 8.5 | 4.4 | | | | | |
| # of observations trimmed | 304 | 744 | | | | | |

Note: This table shows the estimated effects comparing different intensities of musical practice. All outcome variables are standardized to mean zero and variance one. Higher grades are better. The results are obtained by applying the Farrell (2015) estimator using Post-Lasso with penalty chosen at the minimum of 10-fold cross-validated MSE. State and school track dummies enter the selection unpenalized. Standard errors in brackets are clustered at school level. *, **, *** indicate statistical significance at the 10%, 5%, 1% level, respectively.

D.3.4 No state and school track dummies fix in the model

The baseline analysis leaves state and school track dummies unpenalized. This means they are fixed in the models and only additional variables are selected. The idea is to provide the estimator crucial information derived from knowledge about the institutional background saying that states and especially different school tracks might differ substantially. The results in Table D.3.4.1 show that this is not necessary. While the baseline model starts out with 19 dummies included and adds variables, the unpenalized version selects even less than a total of 19 variables. Still the obtained results are remarkably similar to the baseline.

*Table D.3.4.1: Baseline results with no variables fix in the model*

|  | Binary | Multiple | | | | | |
|---|---|---|---|---|---|---|---|
|  | Any - No | Low - No | Med - No | High - No | Med - Low | High - Low | High - Med |
| *Cognitive Skills (standardized)* | | | | | | | |
| Science | 0.11*** | 0.05 | 0.15*** | 0.17*** | 0.10** | 0.12*** | 0.02 |
|  | (0.02) | (0.03) | (0.03) | (0.03) | (0.04) | (0.04) | (0.04) |
| Math | 0.09*** | 0.07** | 0.11*** | 0.13*** | 0.04 | 0.06 | 0.02 |
|  | (0.02) | (0.03) | (0.03) | (0.03) | (0.03) | (0.04) | (0.04) |
| Vocabulary | 0.11*** | 0.02 | 0.15*** | 0.21*** | 0.13*** | 0.19*** | 0.06 |
|  | (0.03) | (0.03) | (0.03) | (0.03) | (0.03) | (0.04) | (0.04) |
| Reading | -0.01 | 0.03 | -0.04 | 0.00 | -0.07 | -0.03 | 0.04 |
|  | (0.03) | (0.04) | (0.03) | (0.04) | (0.04) | (0.04) | (0.04) |
| ICT | 0.14*** | 0.08** | 0.16*** | 0.19*** | 0.08* | 0.11*** | 0.03 |
|  | (0.02) | (0.03) | (0.03) | (0.03) | (0.04) | (0.04) | (0.04) |
| *School performance (standardized)* | | | | | | | |
| German grade | 0.12*** | 0.10*** | 0.19*** | 0.14*** | 0.03 | 0.04 | 0.01 |
|  | (0.03) | (0.04) | (0.03) | (0.04) | (0.04) | (0.05) | (0.04) |
| Math grade | 0.05* | 0.04 | 0.13*** | 0.03 | 0.03 | -0.01 | -0.04 |
|  | (0.03) | (0.04) | (0.03) | (0.04) | (0.04) | (0.05) | (0.05) |
| Average grade German & math | 0.10*** | 0.09** | 0.11*** | 0.09** | 0.03 | 0.00 | -0.03 |
|  | (0.03) | (0.04) | (0.03) | (0.04) | (0.04) | (0.05) | (0.04) |
| *Big Five (standardized)* | | | | | | | |
| Extraversion | 0.03 | 0.00 | 0.04 | 0.05 | 0.04 | 0.05 | 0.01 |
|  | (0.03) | (0.04) | (0.04) | (0.04) | (0.04) | (0.05) | (0.05) |
| Agreeableness | 0.12*** | 0.12*** | 0.11*** | 0.12*** | 0.00 | 0.00 | 0.00 |
|  | (0.03) | (0.04) | (0.04) | (0.04) | (0.04) | (0.05) | (0.05) |
| Conscientiousness | -0.04 | -0.04 | -0.06 | 0.00 | -0.02 | 0.04 | 0.06 |
|  | (0.03) | (0.04) | (0.03) | (0.04) | (0.04) | (0.05) | (0.05) |
| Neuroticism | 0.001 | 0.05 | -0.01 | -0.04 | -0.06 | -0.09* | -0.03 |
|  | (0.03) | (0.04) | (0.04) | (0.04) | (0.04) | (0.05) | (0.05) |
| Openness | 0.30*** | 0.14*** | 0.33*** | 0.49*** | 0.19*** | 0.35*** | 0.16*** |
|  | (0.03) | (0.04) | (0.03) | (0.04) | (0.04) | (0.05) | (0.04) |
| (Mean) # of selected variables for treatment | 15 | | | | 12.3 | | |
| Mean # of selected variables for outcomes | 7.8 | | | | 6.2 | | |
| # of observations trimmed | 11 | | | | 38 | | |

Note: This table shows the estimated effects comparing different intensities of musical practice. All outcome variables are standardized to mean zero and variance one. Higher grades are better. The results are obtained by applying the Farrell (2015) estimator using Post-Lasso with penalty chosen at the minimum of 10-fold cross-validated MSE. Standard errors in brackets are clustered at school level. *, **, *** indicate statistical significance at the 10%, 5%, 1% level, respectively.

D.3.5 Only main effects considered in the analysis

The main analysis considers more than 10,000 covariates including interactions and polynomials to allow for flexible modelling. This sensitivity check investigates whether blowing up number of covariates makes a substantial difference compared to the inclusion of only main effects. Table D.3.5.1 shows the results when only the 328 main effects and 532 school dummies are available for the model selection. The estimates are very similar to the main results. This suggests that the increased flexibility does not change much and main effects are sufficient to provide a good approximation of the underlying functional forms of the treatments and outcomes in this application.

*Table D.3.5.1: Baseline results with only main effects considered*

|  | Binary | Multiple | | | | | |
|---|---|---|---|---|---|---|---|
|  | Any - No | Low - No | Med - No | High - No | Med - Low | High - Low | High - Med |
| *Cognitive Skills (standardized)* | | | | | | | |
| Science | 0.10*** | 0.03 | 0.13*** | 0.18*** | 0.09** | 0.14*** | 0.05 |
|  | (0.02) | (0.03) | (0.03) | (0.03) | (0.04) | (0.04) | (0.04) |
| Math | 0.07*** | 0.05 | 0.09*** | 0.11*** | 0.05 | 0.06 | 0.01 |
|  | (0.02) | (0.03) | (0.03) | (0.03) | (0.03) | (0.04) | (0.04) |
| Vocabulary | 0.11*** | 0.02 | 0.15*** | 0.19*** | 0.13*** | 0.17*** | 0.05 |
|  | (0.02) | (0.03) | (0.03) | (0.03) | (0.03) | (0.04) | (0.04) |
| Reading | -0.03 | 0.00 | -0.04 | -0.01 | -0.05 | -0.01 | 0.03 |
|  | (0.03) | (0.03) | (0.03) | (0.04) | (0.04) | (0.04) | (0.04) |
| ICT | 0.12*** | 0.06** | 0.15*** | 0.19*** | 0.08** | 0.12*** | 0.04 |
|  | (0.02) | (0.03) | (0.03) | (0.03) | (0.03) | (0.04) | (0.04) |
| *School performance (standardized)* | | | | | | | |
| German grade | 0.11*** | 0.09*** | 0.11*** | 0.15*** | 0.02 | 0.06 | 0.05 |
|  | (0.03) | (0.03) | (0.03) | (0.04) | (0.04) | (0.05) | (0.04) |
| Math grade | 0.06** | 0.02 | 0.08** | 0.02 | 0.05 | 0.00 | -0.05 |
|  | (0.03) | (0.04) | (0.04) | (0.04) | (0.04) | (0.05) | (0.05) |
| Average grade German & math | 0.09*** | 0.08** | 0.11*** | 0.10** | 0.03 | 0.02 | -0.01 |
|  | (0.03) | (0.04) | (0.03) | (0.04) | (0.04) | (0.05) | (0.05) |
| *Big Five (standardized)* | | | | | | | |
| Extraversion | 0.03 | 0.03 | 0.04 | 0.05 | 0.02 | 0.03 | 0.01 |
|  | (0.03) | (0.04) | (0.04) | (0.04) | (0.04) | (0.05) | (0.05) |
| Agreeableness | 0.10*** | 0.10*** | 0.09** | 0.11*** | -0.01 | 0.01 | 0.02 |
|  | (0.03) | (0.04) | (0.04) | (0.04) | (0.04) | (0.05) | (0.05) |
| Conscientiousness | -0.03 | -0.03 | -0.06* | -0.02 | -0.02 | 0.02 | 0.04 |
|  | (0.03) | (0.04) | (0.03) | (0.04) | (0.04) | (0.05) | (0.05) |
| Neuroticism | 0.00 | 0.04 | -0.02 | -0.05 | -0.06 | -0.09* | -0.03 |
|  | (0.03) | (0.04) | (0.04) | (0.04) | (0.04) | (0.05) | (0.05) |
| Openness | 0.31*** | 0.15*** | 0.32*** | 0.49*** | 0.17*** | 0.34*** | 0.17*** |
|  | (0.03) | (0.04) | (0.03) | (0.04) | (0.04) | (0.05) | (0.05) |
| (Mean) # of selected variables for treatment | 13 | 6.0 | | | | | |
| Mean # of selected variables for outcomes | 8.4 | 4.2 | | | | | |
| # of observations trimmed | 9 | 46 | | | | | |

Note: This table shows the estimated effects comparing different intensities of musical practice. All outcome variables are standardized to mean zero and variance one. Higher grades are better. The results are obtained by applying the Farrell (2015) estimator using Post-Lasso with penalty chosen at the minimum of 10-fold cross-validated MSE. State and school track dummies enter the selection unpenalized. Standard errors in brackets are clustered at school level. *, **, *** indicate statistical significance at the 10%, 5%, 1% level, respectively.

D.3.6 Random Forest

We apply Post-Lasso throughout the paper to estimate the nuisance parameters. This might be problematic if the interaction terms and polynomials that are supplied to the Post-Lasso algorithm are not sufficient to capture the non-linearities in the conditional expectations of the nuisance parameters. Further, Post-Lasso requires the stated sparsity assumptions that might be critical.

As an alternative, we consider Random Forests (Breiman, 2001) using the R package `randomForest` to estimate the nuisance parameters. We grow 2000 single trees for each forest and select number of variables randomly sampled as candidates at each split, the tuning parameter, that minimizes the out-of-bag error. The nuisance parameters that are plugged into the efficient score are estimated out-of-bag to avoid overfitting.

*Table D.3.6.1: Random Forest is used to estimate nuisance parameters*

|  | Binary | Multiple | | | | | |
|---|---|---|---|---|---|---|---|
|  | Any - No | Low - No | Med - No | High - No | Med - Low | High - Low | High - Med |
| Cognitive Skills (standardised) | | | | | | | |
| Science | 0.10*** | 0.06 | 0.14*** | 0.15*** | 0.08* | 0.10* | 0.01 |
|  | (0.02) | (0.04) | (0.03) | (0.04) | (0.04) | (0.05) | (0.04) |
| Math | 0.07*** | 0.07* | 0.10*** | 0.11*** | 0.03 | 0.05 | 0.01 |
|  | (0.02) | (0.03) | (0.03) | (0.03) | (0.04) | (0.04) | (0.04) |
| Vocabulary | 0.10*** | 0.01 | 0.15*** | 0.19*** | 0.14*** | 0.18*** | 0.04 |
|  | (0.02) | (0.03) | (0.03) | (0.03) | (0.04) | (0.04) | (0.04) |
| Reading | 0.003 | 0.03 | -0.01 | 0.02 | -0.04 | -0.01 | 0.03 |
|  | -0.030 | (0.04) | (0.03) | (0.04) | (0.04) | (0.05) | (0.05) |
| ICT | 0.11*** | 0.06* | 0.15*** | 0.16*** | 0.09** | 0.10** | 0.01 |
|  | (0.02) | (0.03) | (0.03) | (0.04) | (0.04) | (0.04) | (0.04) |
| School performance (standardised) | | | | | | | |
| German grade | 0.12*** | 0.14*** | 0.13*** | 0.13** | -0.01 | -0.003 | 0.003 |
|  | (0.03) | (0.04) | (0.03) | (0.04) | (0.04) | (0.05) | (0.05) |
| Math grade | 0.04 | 0.04 | 0.07* | -0.02 | 0.02 | -0.06 | -0.09* |
|  | (0.03) | (0.04) | (0.04) | (0.04) | (0.04) | (0.05) | (0.05) |
| Average grade German & math | 0.10*** | 0.10*** | 0.12*** | 0.07* | 0.02 | -0.03 | -0.05 |
|  | (0.03) | (0.04) | (0.03) | (0.04) | (0.04) | (0.05) | (0.05) |
| Big Five (standardised) | | | | | | | |
| Extraversion | 0.03 | 0.001 | 0.06* | 0.05 | 0.06 | 0.04 | -0.02 |
|  | (0.03) | (0.04) | (0.04) | (0.05) | (0.05) | (0.05) | (0.05) |
| Agreeableness | 0.10*** | 0.08* | 0.08* | 0.11* | -0.003 | 0.02 | 0.03 |
|  | (0.03) | (0.04) | (0.04) | (0.04) | (0.05) | (0.05) | (0.05) |
| Conscientiousness | -0.03 | -0.03 | -0.05 | 0.01 | -0.03 | 0.03 | 0.06 |
|  | (0.03) | (0.04) | (0.03) | (0.04) | (0.04) | (0.05) | (0.05) |
| Neuroticism | 0.004 | 0.05 | -0.03 | -0.03 | -0.07* | -0.08 | -0.01 |
|  | (0.03) | (0.04) | (0.04) | (0.04) | (0.04) | (0.05) | (0.05) |
| Openness | 0.31*** | 0.12** | 0.34*** | 0.51*** | 0.22*** | 0.39*** | 0.17*** |
|  | (0.03) | (0.04) | (0.03) | (0.04) | (0.04) | (0.05) | (0.05) |

Note: This table shows the estimated effects comparing different intensities of musical practice. All outcome variables are standardized to mean zero and variance one. Higher grades are better. The results are obtained by applying the Farrell (2015) estimator using Random Forest with number of variables considered at each split determined by out-of-bag validation. Standard errors in brackets are clustered at school level. *, **, *** indicate statistical significance at the 10%, 5%, 1% level, respectively.

# Appendix E: Propensity score and common support

## E.1 Propensity score

Table E.1.1 shows the average marginal effects of the additionally selected variables for the propensity scores. They should be interpreted with caution as they were only chosen to optimize prediction. However, it is interesting to see that mainly interactions with gender and the availability of cultural books in households are selected as most predictive for different intensities of music. This suggests that gender and parental tastes are the main drivers into playing music at least after controlling for information contained in state and school track dummies.

*Table E.1.1: Average marginal effects of the additionally selected variables in the propensity score estimations of the baseline*

| | Binary treatment | |
|---|---|---|
| | Average marginal effects | S.E. |
| *Selected variables Any vs. No* | | |
| Biological mother, adoptive mother, foster mother in HH * student female | 0.04 | (0.03) |
| HH size * student female | 0.02*** | (0.01) |
| More than 500 books in HH * books with poems in HH | 0.10*** | (0.02) |
| Desk to study in HH * student female | 0.02 | (0.06) |
| Room just for student in HH * classic literature in HH | 0.03 | (0.03) |
| Room just for student in HH * student female | 0.03 | (0.04) |
| Classic literature in HH * books with poems in HH | 0.08*** | (0.03) |
| Classic literature in HH * works of art in HH | 0.05** | (0.02) |
| Books with poems in HH * student female | 0.03 | (0.02) |
| | Multiple treatment | |
| | Average marginal effects | S.E. |
| *Selected variables No vs. rest (Any)* | | |
| *Like Any vs. No above but with negative coefficients* | | |
| *Selected variables Low vs. rest* | | |
| Books useful for homework in HH * student female | 0.08*** | (0.01) |
| *Selected variables Medium vs. rest* | | |
| HH size * student female | 0.01*** | (0.004) |
| Books with poems in HH * student female | 0.06*** | (0.02) |
| *Selected variables High vs. rest* | | |
| Books with poems in HH * classic literature in HH | 0.03** | (0.01) |
| Assets in HH: savings book * classic literature in HH | 0.03 | (0.02) |
| More than 500 books in HH * mum higher tertiary education | 0.04* | (0.02) |
| More than 500 books in HH * books with poems in HH | 0.03** | (0.02) |
| Parents in favor of gender equality for vocational training * classic literature in HH | 0.10*** | (0.02) |
| Classic literature in HH * books with poems in HH | 0.03* | (0.02) |

Note: The estimation is based on Post-Lasso Logit models using the respective selected additional variables in the baseline analysis as well as school track and state dummies that are omitted due to space and confidentiality reasons. All shown variables are interaction terms. We obtain standard errors (S.E.) from a clustered bootstrap at school level with 4,999 replications. *, **, *** mean statistically different from zero at the 10%, 5%, 1% level, respectively. HH is the abbreviation for household.

## E.2 Common support

We enforce common support by trimming all observations with propensity scores below the largest minimum propensity score in the different treatment groups as well as propensity scores above the smallest maximum propensity score in the different treatment groups. For the binary treatment, Figure E.2.1 shows that common support is not an issue in our application.

*Figure E.2.1: Overlap of the propensity score for the baseline with binary treatment*

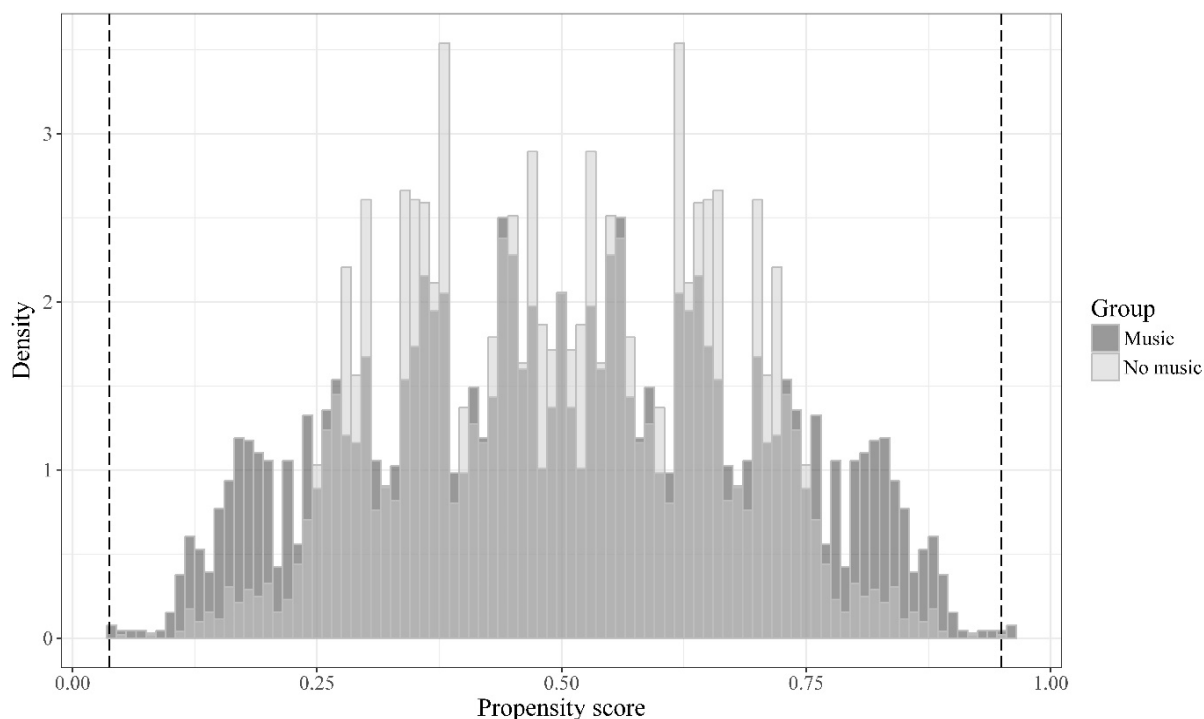

Note: Histogram of propensity score based on Post-Lasso Logit with penalty chosen at the minimum of 10-fold cross-validated MSE. Binwidth 0.01. Dashed lines show the lower and upper threshold of trimming

The illustration of an overlap of four different propensity scores in the multiple treatment setting is too confusing and therefore not shown graphically. However, Table E.2.1 reports the number of observations trimmed in binary and multiple treatment settings. At the baseline only seven and 37 observations are off support and trimmed, respectively. But also for the specification with many more variables obtained from the 2SE+ rule, at most 235 (4%) of the observations are trimmed.

*Table E.2.1: Observations dropped to enforce common support for different penalty term choices*

| # of observations dropped to enforce common support | Dummies | 1SE | **Min** | 1SE+ | 2SE+ |
|---|---|---|---|---|---|
| Binary treatment | 0 | 1 | **7** | 22 | 21 |
| Multiple treatment | 8 | 23 | **39** | 162 | 235 |

    Common support is enforced after prediction of the outcomes. Another possibility would be to trim before predicting the outcomes such that only the outcome is approximated only in "relevant" regions of the covariate space, which could improve efficiency. However, our case where different penalty term choices lead to different samples would require separate outcome predictions for each penalty. This possibility is neglected for computational reasons. An investigation of how and at which point of the estimation procedure common support should be enforced is left for further research.